\documentclass[11pt,a4paper]{article}
\pdfoutput=1
\usepackage{jheppub}
\usepackage{amsfonts}
\usepackage{bbm}
\usepackage{graphicx}
\usepackage{caption}
\usepackage{subcaption}
\usepackage{bigints}
\usepackage[normalem]{ulem}
\usepackage{slashed}

\def\i{\iota}

\def\D{{\rm D}}
\def\d{{\rm d}}

\def\sD{{\slashed{\rm D}}}

\newcommand{\cD}{\mathcal{D}}

\newcommand{\cL}{\mathcal{L}}
\newcommand{\cM}{\mathcal{M}}
\newcommand{\cN}{\mathcal{N}}

\newcommand{\Li}{{\rm Li}}

\newcommand\vG{{\Phi}}
\newcommand\vD{{\omega}}
\newcommand\vR{{\Theta}}
\newcommand\vGh{\hat{\Phi}}

\newcommand{\half}{\frac{1}{2}}
\newcommand{\ndt}{\noindent}

\newcommand{\nn}{\nonumber}

\def\i{\mathrm{i}}

\def\p{\partial}

\def\bea{\begin{eqnarray}}
\def\eea{\end{eqnarray}}
\def\be{\begin{equation}}
\def\ee{\end{equation}}
\def\ba{\begin{array}}
\def\ea{\end{array}}

\newcommand{\bem}{\begin{pmatrix}}
\newcommand{\eem}{\end{pmatrix}}

\def\={\;  = \;}
\def\+{\, + \,}

\def\wt{\widetilde}

\def\bar{\overline}

\def\rt2{\sqrt{2}}

\title{Duality and Transport for Supersymmetric Graphene from the
Hemisphere Partition Function}

\author[a]{\small{Rajesh Kumar Gupta},}
\author[b]{\small{Christopher P.\ Herzog}}
\author[b]{\small{and Imtak Jeon}}
\affiliation[a]{\small{Department of Physics, Indian Institute of Technology Ropar,
Rupnagar, Punjab 140001, India}}
\affiliation[b]{\small{Department of Mathematics, King's College London, The Strand, London WC2R 2LS, UK}}

\emailAdd{rajesh.gupta@iitrpr.ac.in}
\emailAdd{christopher.herzog@kcl.ac.uk}
\emailAdd{imtak.jeon@kcl.ac.uk}
\abstract{
We use localization to compute the partition function of a four dimensional, supersymmetric, 
abelian gauge theory 
on a hemisphere coupled to charged matter on the boundary.
Our theory has eight real supercharges in the bulk of which four are broken by the presence of the boundary.
The main result is that 
the partition function is identical to that of ${\mathcal N}=2$ abelian 
Chern-Simons theory on a three-sphere coupled to chiral multiplets, 
but where the
quantized Chern-Simons level is replaced by an arbitrary complexified gauge coupling $\tau$.
The localization reduces the path integral to a single ordinary integral over a real variable.
This integral in turn allows us to calculate the scaling dimensions of certain protected operators
and two-point functions of abelian symmetry currents at arbitrary values of $\tau$.
Because the underlying theory has conformal symmetry, 
the current two-point functions 
 tell us the zero temperature conductivity of the Lorentzian versions of these
theories at any value of the coupling.
We comment on S-dualities which relate different theories of supersymmetric graphene. We identify a couple of self-dual theories for which the complexified conductivity associated to the U(1) gauge symmetry is $\tau/2$.
}

\begin{document}
\maketitle

\section{Introduction}

The theory of a four dimensional photon interacting with charged matter on a three dimensional boundary
or interface possesses a number of remarkable properties.  Studied a number of years ago in the context
of D-brane physics, graphene, and toy models of confinement \cite{Gorbar:2001qt,Reystalk,Kaplan:2009kr}, 
the model has had an increasing impact on the literature in recent years.  
See 
refs.\ \cite{Grignani:2019zxc,Hsiao:2018fsc,Hsiao:2017lch,Teber:2018jdh,DiPietro:2019hqe,Herzog:2017xha}
for more recent works on this theory.
Our work here is an application of a supersymmetric version \cite{Herzog:2018lqz} of this theory.

We compute the partition function on a four dimensional hemisphere 
$HS^4$ using localization techniques.
While this particular localization calculation has to our knowledge not been performed, 
it relies heavily on earlier results.  
 We take advantage of previous localization calculations on $S^4$ \cite{Pestun:2007rz,Hama:2012bg} and 
 on $S^3$ \cite{Hama:2011ea,Hama:2010av,Kapustin:2009kz}.  There is even a closely related localization calculation for a non-abelian gauge theory on $HS^4$ \cite{Gava:2016oep}.\footnote{%
  See also refs.\ \cite{Dedushenko:2018aox,Dedushenko:2018tgx} for further results on localization in the presence of a boundary.
 }
   Ref.\ \cite{Gava:2016oep} is in one respect more general 
 than our case, but in another not general enough.   
 While ref.\ \cite{Gava:2016oep} focuses on nonabelian gauge theory, our theory is abelian.
  While ref.\ \cite{Gava:2016oep} restricts to Neumann or Dirichlet conditions for the gauge field, 
the novel ingredient in our work 
 is that we allow dynamical degrees of freedom 
 on the boundary to couple to the bulk gauge field.  In particular, we allow for $n_+$ and $n_-$ oppositely charged 
 chiral multiplets.

One remarkable feature of this graphene-like theory is that the complexified gauge coupling,
\be
\tau \equiv \frac{2 \pi \i }{g^2} + \frac{\theta}{2\pi}  \, ,
\ee
 is exactly marginal
\cite{Teber:2018jdh,Herzog:2017xha,Dudal:2018pta,Herzog:2018lqz}.  Thus we have an example of a boundary (or interface) conformal field theory for all values of $\tau$.  Localization provides a method for determining the dependence of the partition function on $\tau$ and studying precisely how the theory may change as a function of this coupling.  
As emphasized in ref.\ \cite{DiPietro:2019hqe}, 
at certain special values of the complexified coupling, associated with cusps in the
action of $SL(2, {\mathbb Z})$ on $\tau$, the boundary theory ``decouples'' from the bulk and becomes three dimensional. 
At these values, the complexified coupling $\tau$ is reinterpreted as a quantized -- either integer or half-integer -- 
Chern-Simons level $k$.  Our partition function reduces to that of abelian Chern-Simons theory on a three sphere.
Indeed, our main result is that our hemisphere partition function is identical to the three sphere ${\mathcal N}=2$ 
abelian Chern-Simons theory partition
function (with matter) with the replacement of $\tau$ for $k$.   

Our hemisphere partition function (\ref{Z}) takes the form of a single definite integral over a function involving dilogarithms.  
In addition to the complexified coupling $\tau$, the integral depends on a number of potentials $q_i$ which control the mixing of the R-symmetry with other U(1) symmetries in the theory.  In analogy with the arguments for 3d Chern-Simons theories \cite{Jafferis:2010un,Closset:2012vg}, we expect conformal symmetry picks out special values $q_i^*$ 
which minimize the absolute value of the partition function $|Z|$.  The vanishing of $\partial_{q_i} |Z|$ correlates with the 
vanishing of certain one point functions of boundary operators, while the positivity of $\partial_{q_i} \partial_{q_j} |Z|$
means certain boundary two point functions are reflection positive.

While we cannot evaluate the partition function 
integral in general, we can analyze it in a variety of limits.  We use the saddle point approximation to extract results both at large $|\tau|$ and large $n$.  
In the decoupling limit $\tau = k$, 
we use contour integration to reduce the integral to a sum over residues \cite{Garoufalidis:2014ifa}. 
We also use this same contour integration technique to perform an S duality on our integral, $\tau \to - 1 / \tau$.
We identify a couple of theories which have self dual points.

Interestingly, we are able to compute quantities associated with transport in our theory.  
A second derivative of our partition function with respect to the $q_i$ gives us two-point functions
of U(1) currents \cite{Closset:2012vg}.  As our theories are conformal, we can map these
current-current two-point functions from the sphere back to flat space.  
A Kubo formula then relates these two-point functions
to associated conductivities.  Through self duality, we can also establish the conductivity
with respect to the U(1) gauge symmetry has a particularly simple form at self dual points, namely 
$2 \pi (\sigma_H + \i \sigma) = \tau/2$
in our two examples.  
In the condensed matter context, this ``quantum conductivity''
has been associated with the conductivity at quantum critical points, for example the conductivity of a metallic thin film
at a superfluid insulator transition \cite{Fisher:1990zza,Wen:1990gi} or the conductivity of a quantum Hall system at the transition between two plateaux \cite{Engel:1993zz,Shahar:1997zz}.  Note in our system, we have access to
the conductivities at arbitrary coupling, not just perturbatively where there is expected to be a quasiparticle interpretation.
Additionally, we know precisely what the field theory interpretation is, unlike for example in certain bottom-up holographic models in AdS/CFT.

An outline of the manuscript follows.  In section \ref{sec:bc}, we discuss supersymmetry preserving boundary conditions
in our Euclidean framework.  In section \ref{sec:ks}, we review Killing spinors on $S^4$.  Section \ref{sec:action} contains
the action for our graphene-like theory.  Section \ref{sec:localization} presents the localization calculation and introduces
two techniques to analyze the resulting integral: saddle point approximation and a contour integration.  
Finally, section \ref{sec:sduality} presents the results on S duality and transport.
We conclude with a brief discussion in section \ref{sec:discussion}. 
Appendix \ref{app:conventions} presents our conventions for fermions while appendix \ref{app:specialcases} gives some results for the partition function in simple special cases.
Readers not interested in the details of the localization calculation may wish to stare briefly at the definition of the theory
 (\ref{Vaction}), (\ref{FI}), and (\ref{ChiralAction}) before proceeding directly to section \ref{sec:localization}.

\section{Supersymmetry and Boundary Conditions}
\label{sec:bc}

Two facts about supersymmetry greatly constrain our problem.  The first is that most of the localization technology
available requires the existence of a continuous R-symmetry.  The second is that the presence of a boundary 
will break at least half of the supersymmetry present in the bulk, because supercharges square to translation generators.
The simplest theory we may consider in 4d should thus have at least ${\mathcal N}=2$ supersymmetry with 
(in more familiar Lorentzian signature) $SU(2) \times U(1)$ R-symmetry.  The boundary will have 3d ${\mathcal N}=2$ supersymmetry and 
preserve only a diagonal $U(1) \subset SU(2)$ of the bulk R-symmetry.  

Let us see how this breaking works at the level of the supersymmetry algebra.  For 4d ${\mathcal N}=2$ SUSY, it is convenient to introduce symplectic Majorana fermions which have a charge conjugation condition
\be
\label{symplecticdef}
\bar Q_i = \varepsilon_{ij} Q^{jT} C \ ,
\ee
where $\gamma^{\mu T} = C \gamma^\mu C^{-1}$.  (See appendix \ref{app:conventions} for our conventions.)
The relevant part of the ${\mathcal N}=2$ algebra is given by
\be\label{QQalg}
\{ Q^i , \bar Q_j \} = 2 \i \delta^i_j \sD \,,
\ee
where $iD_{\mu}$ generates translation in 4 dimensions.
The $Q_i$ transform as a doublet under the $SU(2)$ R-symmetry and via $Q_i \to e^{\eta \gamma^5} Q_i$ under the additional $U(1)$.  As we work in Euclidean signature, $\eta$ is a real parameter, and the symmetry is not a compact $U(1)$ but actually $SO(1,1)$.  

The presence of a boundary breaks translation invariance in the normal direction.  We are thus looking for a subalgebra which generates only translations along the boundary, $\i {\rm D}_A$.  We can define this subalgebra through projectors $\Pi_\pm{}^i{}_{j}$ that
preserve the tangential gamma matrices \cite{Herzog:2018lqz}:
\be
\label{projcondition}
{\Pi_+}^i{}_k \gamma^\mu {\bar \Pi_+}^k{}_{j} = \delta^\mu_A {\Pi_+}^i{}_{j} \gamma^A \ . 
\ee  
The definition of $\bar \Pi_\pm$ is induced from the definition of the symplectic Majorana condition (\ref{symplecticdef}):
\be
{{\bar \Pi_\pm}^j}_i  = - C^{-1} \varepsilon^{jk} {(\Pi^T_\pm)_k}^l \varepsilon_{li} C \ .
\ee
The projectors satisfy the usual properties $\Pi_+ + \Pi_- = 1$ and $\Pi_+ \Pi_- = 0$. 
From (\ref{projcondition}) follow the relations
\be
\label{Pigammacomm}
{\bar \Pi_\pm}^i{}_{j} \gamma^A = \gamma^A {\Pi_\pm}^i{}_{j} \ , \; \; \;
{\bar \Pi_\pm}^i{}_{j} \gamma^n = \gamma^n {\Pi_\mp}^i{}_{j} \ , \; \; \; 
{\Pi_\pm}^i{}_{j} \gamma^5 = \gamma^5 {\Pi_\mp}^i{}_{j} \ .
\ee

Projectors satisfying our conditions are
\be\label{projector}
{\Pi_\pm}^i_{j} = \frac{1}{2} ( \delta^i_{j} \pm \i \vec v \cdot {\vec \tau}^i{}_{j} \, \gamma^5 \gamma^n e^{\eta \gamma^5}) \ .
\ee
The vector $\vec \tau = (\sigma_1, \sigma_2, \sigma_3)$ of Pauli spin matrices generates the $SU(2)$ R-symmetry,
while $\vec v$ is a unit vector which determines how the $SU(2)$ is broken down to a $U(1)$ subgroup by the boundary.
The $\eta$ parameter similarly determines how the  bulk $SO(1,1)$ R-symmetry is broken. 
The 3d supersymmetry algebra we take to be generated by ${\Pi_+}^i{}_{j} Q^j$.  

One caveat to obtain this projection is that the analysis is not for the general curved space but for the flat space. As the projector $\Pi_\pm{}^{i}_{j}$ is not covariantly constant, the projected generator ${\Pi_+}^i{}_{j} Q^j$ is not in general a symmetry generator or, equivalently, the corresponding projected spinor ${\Pi_+}^i{}_{j} \xi^j$ is not the Killing spinor. Thus the algebra \eqref{QQalg} will not be simply projected to its tangential direction . However, at the boundary of our example $HS^4$,  the projector will be covariantly constant as there is no spin connection along the normal direction and  no mixing between normal and tangential directions. The projected spinor at the boundary is the $3d$ Killing spinor, projecting the algebra \eqref{QQalg} to its tangential direction.  Therefore, we can use the projector  \eqref{projector} for our example. 

For simplicity, in what follows we take $\vec v = (0,0,1)$ and $\eta = 0$. As ${\Pi_\pm}^i_{j}$ becomes diagonal in the SU(2) basis, it is convenient to identify projectors associated with each basis element:
\be\label{projector2}
{(\Pi_+)^1}_1 = \pi_+ \ , \; \; \; {(\Pi_+)^2}_2 = \pi_- \  \; \; \; \mbox{where   } \pi_\pm = \frac{1}{2} (1 \pm \i \gamma^5 \gamma^n) \ .
\ee
Another simplification is that ${\bar \Pi_\pm}^i{}_{j} ={ \Pi_\pm}^i_{j}$.  Because the $\gamma^A$ commute with ${\Pi_\pm}^i_{j}$ (\ref{Pigammacomm}),  we can use the tangential 4d gamma matrices $\gamma^A$
to generate our 3d gamma matrix algebra:
\be
\Gamma^A = \pi_+ \gamma^A \ .
\ee 
The projection ${\Pi_+}^i_{j} Q^j$ can be broken down into components
\be
{(\Pi_+)^1}_1 Q^1 = \pi_+ Q^1 \equiv Q_+ \ , \; \; \; {(\Pi_+)^2}_2 Q^2 = \pi_- Q^2 \equiv \gamma^n Q_+' \ .
\ee

The charge conjugation matrix in 3d needs to be altered slightly from its 4d version:
\be
\widetilde C = C \gamma^n \ .
\ee
The extra $\gamma^n$ ensures 
that $\widetilde C \pi_\pm = \pi_\pm^T \widetilde C$.  
The alteration leads to some sign differences in 3d, e.g.\
$\widetilde C \Gamma^A \widetilde C^{-1} = - (\Gamma^A)^T$.  
With the new charge conjugation matrix $\widetilde C$ in hand,
we define the 3d barred spinors as
\be
\widetilde Q_+ = Q_+^T \widetilde C \ .
\ee
The 4d symplectic Majorana condition (\ref{symplecticdef}) implies a 3d symplectic Majorana condition 
\be
Q_+^\dagger = \widetilde Q'_+ \ , \; \; \; Q'^\dagger_+ = - \widetilde Q_+ \ .
\ee

\section{Killing Spinors on Spheres}
\label{sec:ks}

On a curved manifold, the supersymmetry is generated by Killing spinors.  One of our first chores is then to work
out the Killing spinors on $HS^4$ and determine which ones are compatible with the boundary conditions
imposed by the projector $\Pi_+$.  

We can write a metric on $S^4$,
\be
\d s^2 = \frac{\sin^2 r}{4} \left[ \sin^2 \theta \, \d \phi^2+ \d \theta^2 + (\d \psi + \cos \theta \, \d \phi)^2 \right] + \d r^2 \ .
\ee
The hemisphere $HS^4$  comes from restricting the range of the polar angle to $0 \leq r \leq \frac{\pi}{2}$.  
We introduce the vielbeins
\begin{eqnarray}
e^\phi &=& \frac{\sin r}{2} \sin \theta \, \d \phi \ , \; \; \;
e^\theta = \frac{\sin r}{2} \d \theta \ ,  \nonumber \\
e^\psi &=& \frac{\sin r}{2} (\d \psi + \cos \theta \, \d \phi) \ , \; \; \;
e^r = \d r \ .
\end{eqnarray}
where we are implicitly assuming an ordering of the gamma matrices, associating
$\gamma^1$ with $e^\phi$, $\gamma^2$ with $e^\theta$, $\gamma^3$ with $e^\psi$
and most importantly $\gamma^4$ or $\gamma^n$ with $e^r$.  
These vielbeins give the following connection one form:
\begin{align}\label{spinconnection}
{\omega^\psi}_r &= \cot r \, e^\psi \ , \nonumber  \\
{\omega^\theta}_r &= \cot r \, e^\theta \ , 
& {\omega^\theta}_\psi &= \frac{1}{\sin r} e^\phi 
\\
{\omega^\phi}_r &= \cot r \, e^\phi \ ,  
& {\omega^\phi}_\psi &=- \frac{1}{\sin r} e^\theta \ , 
& {\omega^\phi}_\theta &= \frac{1}{\sin r} \left( 2\cot \theta \, e^\phi - e^\psi \right) \ . \nonumber
\end{align}

We look for solutions of the Killing spinor equations\footnote{%
 As we work on a sphere, we do not need the full machinery of off-shell supergravity
 to generate the appropriate Killing spinor equation.  We can get away with
 the simple choice here.
 }
\be
\D_\mu \psi = \pm  \frac{\i}{2}  \gamma_\mu \psi \ .
\ee
There are eight solutions
\begin{align}
\psi^{(1)}_\pm &= \left(
\begin{array}{c}
\cos \frac{r}{2} \\
0 \\
\pm \i \sin \frac{r}{2} \\
0
\end{array}
\right) e^{\i \psi/2}
 \ , 
 \; \; \;
&\psi^{(2)}_\pm
&= \left( \begin{array}{c}
0 \\
\cos \frac{r}{2} \\
0\\
\pm \i \sin \frac{r}{2}
\end{array}
\right) e^{-\i \psi/2} \ ,
\\
\chi^{(1)}_\pm &=\left( \begin{array}{c}
\sin \frac{r}{2} \sin \frac{\theta}{2} \\
- \sin \frac{r}{2} \cos \frac{\theta}{2} \\
\mp \i \cos \frac{r}{2} \sin \frac{\theta}{2} \\
\pm \i \cos \frac{r}{2} \cos \frac{\theta}{2}
\end{array}
\right) e^{\i \phi/2} \ , \; \; \;
&\chi^{(2)}_\pm &=\left( \begin{array}{c}
\sin \frac{r}{2} \cos \frac{\theta}{2} \\
\sin \frac{r}{2} \sin \frac{\theta}{2} \\
\mp \i \cos \frac{r}{2} \cos \frac{\theta}{2} \\
\mp \i \cos \frac{r}{2} \sin \frac{\theta}{2}
\end{array}
\right) e^{-\i \phi/2} \ .
\end{align}
The spinors satisfy the following reality conditions
\be
\psi_\mp^{(2)} = C \psi_{\pm}^{(1)*} \ , \; \; \;
\chi_\mp^{(2)} = C \chi_{\pm}^{(1)*} \,,
\ee
which allows us to repackage them into eight sets of symplectic Majorana spinors $\xi^i_a $ as 
\be\ba{ll}\label{8setKS}
\xi_1^i\= \bigl( \chi^{(1)}_+\,,\chi^{(2)}_- \bigr)\,,~~& \xi_2^i\= \bigl( \i\chi^{(1)}_+\,,-\i\chi^{(2)}_- \bigr)\,,
\\
\xi_3^i\= \bigl(\i \chi^{(2)}_+\,,\i \chi^{(1)}_- \bigr)\,,~~& \xi_4^i\= \bigl( \chi^{(2)}_+\,,-\chi^{(1)}_- \bigr)\,,
\\
\xi_5^i\= \bigl( \psi^{(1)}_+\,,\psi^{(2)}_- \bigr)\,,~~& \xi_6^i\= \bigl( \i\psi^{(1)}_+\,,-\i\psi^{(2)}_- \bigr)\,,
\\
\xi_7^i\= \bigl(\i \psi^{(2)}_+\,, \i \psi^{(1)}_- \bigr)\,,~~~~&\xi_8^i\= \bigl( \psi^{(2)}_+\,, - \psi^{(1)}_- \bigr)\,,
\ea\ee
that are orthonormal to each other as $\overline{\xi_{a i}}\xi^i_b = 2\delta_{ab}$. 
The Killing spinor condition on these symplectic Majorana fermions becomes
\be\label{KSEmarojana}
\D_\mu \xi^i_a =  \frac{\i}{2} \gamma_\mu {(\tau_3)^i}_j \xi^j_a \ .
\ee

The spinors $\xi^i_a\,, a=1,2,3,4$ generate a Killing vector in the $\phi$ direction while the spinors $\xi^i_a\,,a=5,6,7,8$ generate a Killing vector in the $\psi$ direction, 
\be\ba{l}
\overline{\xi_{ai}}\gamma^\mu \xi^i_a = (\pm 4\,,0\,, 0\,,0)\,,~~~\mbox{+ for } a=1,2~~~ - \mbox{  for }a=3,4 ,
\\
\overline{\xi_{ai}}\gamma^\mu \xi^i_a = (0\,,0\,, \pm 4\,,0)\,,~~~\mbox{+ for } a=5,6~~~ - \mbox{  for }a=7,8 \,.
\ea\ee

At the equator $r = \frac{\pi}{2}$, the Killing spinors satisfy the projection conditions
\begin{eqnarray}\label{twoprojection1}
\Pi_- \xi_{a}^{i}|_{r=\pi/2} = 0 \ ,&& \; \; \;\mbox{for  }a=1,2,3,4 \, ,
\\
\label{twoprojection2}
\Pi_+ \xi_{a}^{i} |_{r=\pi/2}= 0 \ ,&& \; \; \;\mbox{for  }a=5,6,7,8   \,. 
\end{eqnarray}
We can take the preserved supersymmetry to satisfy one of the two conditions~\eqref{twoprojection1} and~\eqref{twoprojection2}.
On the three sphere, the Killing vectors $v^\mu = (1,0,0)$ for the former condition~\eqref{twoprojection1} and $(0,0,1)$ for the latter condition~\eqref{twoprojection2} are both divergenceless, $\D_\mu v^\mu = 0$. While $(1,0,0)$ satisfies $\D_\mu v_\nu = -\epsilon_{\mu\nu\rho} v^\rho$,   there is a flip in sign for $(0,0,1)$:
$\D_\mu v_\nu = \epsilon_{\mu\nu\rho} v^\rho$.

The condition of the Killing spinors~\eqref{twoprojection1} (or~\eqref{twoprojection2}) reduces  the 4-dimensional Killing spinor equation \eqref{KSEmarojana} to the 3-dimensional Killing spinor equation,
\be
\D^{(3d)}_A \xi_a^i = \frac{\i}{2} (\tau_3)^i{}_{j}\gamma_A \xi^j_a\,,
\ee
at the boundary of the $HS_4$  where $r=\frac{\pi}{2}$.
This reduction can be done  because the mixing components between the normal and tangential directions of boundary  in the spin connections  are absent at the boundary; in fact, from \eqref{spinconnection} we see that
$
{\omega^\psi}_r =  {\omega^\theta}_r ={\omega^\phi}_r =0\,
$  at $r=\frac{\pi}{2}$.

\section{The Action}
\label{sec:action}

\paragraph{Vector Multiplet on an $HS^4$}
~\\
We consider  the abelian vector multiplet on $HS^4$. 
In the bulk, the supersymmetry is generated by the $SU(2)$ pair of Killing spinors~$(\xi^1\,,\xi^2)$,\footnote{%
Adapted from the appendix of ref.\ \cite{Jeon:2018kec}, with field redefinitions for scalars $X= \frac{\i}{2} (S+P)\,,\bar{X}=  \frac{\i}{2} (S-P)$ and the auxiliary field $Y^{ik}\epsilon_{kj} = -\i \vec{D}\cdot \vec{\tau}^i{}_{j}$.
}
\begin{eqnarray}
\label{susyVec}
\delta S&=& \bar\xi_{i}\,\lambda^{i}\,, \nn\\
\delta P&=&{\bar\xi}_{i}\gamma_5{\lambda}^{i}\,,\nn\\
\delta A_{\mu}&=& {\bar\xi}_{i}\gamma_{\mu}\lambda^{i}\,,\\
\delta \vec{D}&=&\i \vec{\tau}^i{}_{j} \overline{\xi_i} \gamma^a \cD_a \lambda^j \nn \, , \\
\delta \lambda^i &=&  \gamma^{a}\partial_{a} S\,{\xi}^{i}- \gamma^{a}\partial_{a} P\,\gamma_5{\xi}^{i}-\half F_{ab}\gamma^{ab}\xi^{i}-\i \vec{D}\cdot \vec{\tau}^{i}{}_{j}\xi^{j} +\half S\,\slashed{\cD}\xi^{i}-\half P\,\slashed{\cD}\gamma_5\xi^{i}\,.\nn
\end{eqnarray}
The supersymmetry algebra \eqref{susyVec} squares to yield translation generators ${\mathcal L}_v$ along with other symmetry generators of the theory\footnote{%
We are using the Grassman even Killing spinors.}
\be\label{algebra1}
\delta^2=\cL_v+\text{Gauge}(\vGh)+\text{Lorentz}(L^{ab})+\text{Scale}\left(\vD\right)+\text{R}_{SO(1,1)}(\vR)+\text{R}_{SU(2)}(\vec{\vR} )\,,
\ee
where
\begin{eqnarray}\label{parameters2}
&&v^{\mu}\= {\bar\xi}_{i}\gamma^{\mu}\xi^i\,,~~
\vGh \= {\bar\xi}_{i}{\xi}^{i}\,S  -  \bar\xi_{ i}\gamma_5\xi^{i}\,P - v^\mu A_{\mu}\,,~~
L_{ab}\= \cD_{[a}v_{b]}+v^\mu \omega_{\mu ab} \, , \\
&&
\vD
 \= \frac{1}{4} \cD_a v^a\,,~~~
\vR\=
-\half \bar{\xi}_i \gamma_5 \slashed{\cD}\xi^i\,,~~~
\vec{\vR} \= -\i \vec{\tau}^j{}_{i}\overline{\xi_j}\slashed{\cD}\xi^i\,. 
\nonumber 
\end{eqnarray}
Explicitly, 
\be\ba{lll}
\delta^2 S&=&  v^\mu \partial_\mu S +\vD S -\vR P \, , \\
\delta^2 P&=& v^\mu \partial_\mu P -\vR  S+\vD P \, , \\
\delta^2 A_\mu &=& v^\nu F_{\nu\mu} + \partial_\mu \vG \, , \\
\delta^2 \lambda^i &=& v^\mu \cD_\mu \lambda^i + \frac{3}{2}\vD \lambda^i -\half \vR \gamma_5\lambda^i +\frac{1}{4}\cD_{[a}v_{b]} \gamma^{ab}\lambda^i + \frac{1}{2}\vec{\Theta}\cdot \vec{\tau}^i{}_{j}\lambda^j \, , \\
\delta^2 D_I  &=&  v^\mu \cD_\mu D_I  + 2 \vD D_I  +  \epsilon_{IJK} \Theta_J D_K \,.
\ea\ee

The presence of the boundary of $HS^4$ breaks half of the supersymmetries.  The supersymmetries are now generated by the four Killing spinors that satisfy the condition~ \eqref{twoprojection1} in terms of the projector introduced in \eqref{projector}. 
The supersymmetric action under this condition is as follows.  It is divided into bulk and  boundary contributions associated with the gauge coupling $g$ and theta angle $\theta$:
\be\label{Vaction}
I_V \=  I_g + I_{\partial g}+ I_{\theta}
\ee
\begin{eqnarray}\label{Sg}
I_{g}&=&\frac{1}{2g^2}\int_{\cM} d^4x \left( \frac{1}{2} F_{\mu\nu}F^{\mu\nu}+ \overline{\lambda_i}\slashed{\cD} \lambda^i + (\partial_\mu P)^2- (\partial_\mu S)^2 - \vec{D}^2 +\frac{R}{6}(P^2 - S^2)\right),
\\
\label{Sbg}
I_{\partial g}&=& \int_{\partial \cM} d^3 x \left( -\frac{\i}{4}\tau_3{}^i{}_{j}\, \overline{\lambda_{i}}\gamma_5 \lambda^j - P (D_3 +\partial_n P)\right)\,,
\\
\label{Stheta}
I_\theta &=& \frac{ \theta}{4\pi^2} \left[ \frac{\i}{4} \int_{\cM} d^4 x F_{\mu\nu}\widetilde{F}^{\mu\nu} -\int_{\partial \cM}d^3 x\left( \frac{1}{2} \tau_3{}^i{}_{j}\,\overline{\lambda_+}_i \lambda^j_+  - \i S(D_3 +\partial_n P )   \right)\right]\,.
\end{eqnarray}
where the masses of scalars are set by conformal coupling with the curvature of $HS^4$ which is taken to be $R=12$. The dual of the field strength is $\widetilde{F}^{\mu\nu}= \half \epsilon^{\mu\nu\lambda\rho}F_{\lambda\rho}$. The theta angle action \eqref{Stheta} includes the projected component of  the gaugino $\lambda^i_+ = \Pi_+{}^i{}_{j} \lambda^j$ .

One can check that the boundary term from the variation of the bulk action \eqref{Sg}  is canceled by the variation of the boundary action \eqref{Sbg} under the condition  \eqref{twoprojection1}. And the action for the theta angle \eqref{Stheta} is supersymmetric by itself.\footnote{%
The action is supersymmetric under the condition $\Pi_+^i{}_{j}\xi^j|_{r=\pi/2}=0$, if we choose 
\be I_{\partial g}= \int_{\partial \cM} d^3 x \left( \frac{\i}{4}\tau_3{}^i{}_{j}\, \overline{\lambda}_{i}\gamma_5 \lambda^j + P (D_3 -\partial_n P)\right)\,.\nn\ee}
Since we are treating a $U(1)$ gauge group, we can also add the Fayet-Iliopoulos type action,
\begin{eqnarray}
\label{FI}
I_{FI}& =&\int_{\cM}d^4x\,\left( \eta (2S + D_3 + \partial_n P)\right)\,.
\end{eqnarray}
The supersymmetry variation of $2S$ is canceled by the variation of $D_3$ using \eqref{KSEmarojana}, and the remaining boundary term is canceled by the variation of $\partial_n P$ using the projection condition~\eqref{twoprojection1}. 

The Euclidean 4-dimensional space allows the symplectic Majorana condition \eqref{symplecticdef} on the spinors,
giving $8$ real degrees of freedom for  fermions and supercharges. 
The action and supersymmetry algebra
are compatible with the supersymmetry transformation rules~\eqref{susyVec} if bosonic fields  satisfy the following reality conditions
\be\label{realitySUSY}
S^\ast= S\,,~~~~P^\ast =P\,,~~~\vec{D}^\ast =\vec{D}\,,~~~A_\mu^\ast = A_\mu\,.
\ee 
However,  for the quantum theory to be well defined and for the real part of the action \eqref{Vaction} to be positive definite, we need to choose
a contour in the path integral where 
\be\label{realityPI}
S^\ast= -S\,,~~~~P^\ast =P\,,~~~\vec{D}^\ast =-\vec{D}\,,~~~A_\mu^\ast = A_\mu\,.
\ee
This condition guarantees \eqref{Sg} is positive definite and that \eqref{Sbg} and \eqref{Stheta} are pure imaginary. 
Along this contour \eqref{realityPI}, the symplectic Majorana condition for fermions is lost. The $SU(2)$ pair of spinors are now two independent spinors, where each one has its own contour for integration. That is to say that we formally double the spinor space in Euclidean 4-dimensional space.

\paragraph{Supersymmetry at boundary}~\\
The boundary value of the four Killing spinors satisfying~\eqref{twoprojection1} defines the 3-dimensional Killing spinors. They are equivalent to the boundary value of the projected spinors $\Pi_+\xi^i$. We define the 3-dimensional boundary Killing spinors by  using  the simple choice of the projector \eqref{projector2}, 
\be\ba{ll}\label{def3dspinor}
\Pi_+^1{}_{1}\xi^1 |_{r=\frac{\pi}{2}}=  \pi_+ \xi^1|_{r=\frac{\pi}{2}} \equiv \zeta_+\,,~~~~~~~~&\Pi_+^2{}_{2}\xi^2|_{r=\frac{\pi}{2}}  = \pi_- \xi^2 |_{r=\frac{\pi}{2}} \equiv \gamma_n \zeta'_+\,.
\ea\ee
The 4-dimensional symplectic Majorana condition \eqref{symplecticdef} leads to the 3-dimensional reality condition
\be\ba{ll}\label{3dSM}
\zeta_+^\dagger =\widetilde{{\zeta'}_+}\equiv {\zeta'_+}^T \widetilde{C}\,,~~~~~~~~&{\zeta'}_+^\dagger =-\widetilde{\zeta_+}\equiv \zeta_+^T \widetilde{C}\,.
\ea\ee
However, we will not impose a reality condition and instead let $\zeta_+$ and $\zeta'_+$ be independent spinors.  For
the path integral, we do not use the reality \eqref{realitySUSY}, but instead \eqref{realityPI}. In other words, we formally double the spinor space in Euclidean three dimensional space.

When we consider this projected supersymmetry at the boundary, the four dimensional $\cN=2$ algebra is reduced to the three dimensional $\cN=2$ algebra
\be\label{algebra3d}
\delta_+^2=\cL_v+\text{Gauge}(\vGh)+\text{Lorentz}(L^{ab})+\text{Scale}\left(\vD\right)+\text{R}_{U(1)}(\vR_{3} )\,,
\ee
where the parameters \eqref{parameters2}
 are reduced to
\begin{eqnarray}
v^A&=&2  \,\widetilde{\zeta'_+}\Gamma^A \zeta_+ =2 \, \widetilde{\zeta_+}\Gamma^A \zeta'_+\,,~~~~~v^n=0\,, \nonumber \\
\hat{\Phi}&=&2\, \widetilde{\zeta'_+}\zeta_+ \,S -v^A A_A
\,,\nonumber \\
\omega&=&\frac{2}{3}\widetilde{\zeta'_+}\Gamma^A \cD_A \zeta_+ +\frac{2}{3}\widetilde{\zeta_+}\Gamma^A \cD_A \zeta'_+ =  \frac{1}{3}\cD_A v^A  \,,\\
\vR_3&=& -\i \frac{4}{3}\widetilde{\zeta'_+}\Gamma^A \cD_A \zeta_+ +\i\frac{4}{3}\widetilde{\zeta_+}\Gamma^A \cD_A \zeta'_+ \ , \nonumber \\
\vR&\=& \vR_1\=\vR_2\=0\,. \nonumber
\end{eqnarray}
Note that the $SU(2)_R\times SO(1,1)_R$ R-symmetry is broken to the diagonal $U(1)_R$ of $SU(2)_R$.   For the choice of the Killing spinors  $\xi_1^i$ in \eqref{8setKS}, these parameters are
\be
v^A \partial_A = 4\partial_\psi\,,
~~~\hat{\Phi}= 2 S- 4 A_\psi\,,~~~ \omega=0\,,~~~\vR_3= 4\,.
\ee
In the absence of charged matter on the boundary, we may impose Dirichlet boundary conditions on $(P, A_n,\lambda^i_-, D_1, D_2)$ and Neumann boundary conditions on $(S, A_A,  \lambda^i_+, D_3)$ of the 4d vector multiplet fields. The nonzero fields $(S, A_A, \lambda_+, \lambda'_+, (D_3+\partial_n P))$ form a 3d vector multiplet at the boundary of $HS^4$, where the $\lambda_+$ and $\lambda'_+$ are defined in the same manner as the Killing spinor \eqref{def3dspinor}.  The supersymmetry transformation of this 3d vector multiplet takes the usual form
\begin{eqnarray}
\delta_+ A_{A}&=&  \widetilde{\zeta'_+}\Gamma_A \lambda_+ +\widetilde{{\zeta_+}}\Gamma_A \lambda'_+ \, ,  \nonumber \\
\delta_+ \lambda_+ &=& -\frac{1}{2}F_{AB}\Gamma^{AB}\zeta_+  -\i (D_3 +\partial_n P)\zeta_+  + \Gamma^A \partial_A S \zeta_+ +\frac{2}{3} S\Gamma^A \cD_A \zeta_+ \, ,  \nonumber \\
\delta_+ \lambda'_+ &=& -\frac{1}{2}F_{AB}\Gamma^{AB}\zeta'_+  +\i (D_3 +\partial_n P)\zeta'_+  - \Gamma^A \partial_A S \zeta'_+ -\frac{2}{3}S\Gamma^A \cD_A \zeta'_+\, , \\
\delta_+ (D_3 +\partial_n P)&=& \i \widetilde{\zeta'_+} \Gamma^A \cD_A \lambda_+  -\i \widetilde{\zeta_+}\Gamma^A \cD_A \lambda_+' +\i \frac{1}{3} \widetilde{\cD_A \zeta'_+} \Gamma^A \lambda_+  -\i \frac{1}{3}\widetilde{\cD_A \zeta_+} \Gamma^A \lambda'_+\, , \nonumber \\
\delta_+ S&=& \widetilde{\zeta'_+}\lambda_+  - \widetilde{\zeta_+}\lambda'_+ \, . \nonumber
\end{eqnarray}
Our next task, adding charged matter on the boundary, leads to a modification of these boundary conditions.

\paragraph{Chiral Multiplet on $S^3$}
~\\
Next, we couple the boundary value of the vector multiplet fields with degrees of freedom living at the boundary $S^{3}$ (located at $r=\frac{\pi}{2}$). The boundary degrees of freedom consist of $n_+ + n_-$ chiral multiplets, $n_+$ with positive gauge charge and $n_-$ with negative gauge charge.  (Sometimes we also set $n_+ = n_- = n$ to get $n$ hypermultiplets.)
These chiral multiplet fields couple to the boundary modes of the bulk $U(1)$ vector multiplet.  Let one of these
chiral multiplets have $U(1)_{R}$ R-charge $q$. The boundary supersymmetry transformations are generated by supersymmetry parameter $\zeta^{i}_{+}$ which satisfy the algebra \eqref{algebra3d}.
This algebra can be realized on the chiral multiplet with the following transformations
\bea
&&\delta_{+}\phi=-2i\wt\zeta'_{+}\psi_{+},\quad\delta_{+}\bar\phi=-2i\wt\zeta_{+}\psi'_{+} \, , \nn\\
&&\delta_{+}\psi_{+}=\i\Gamma^{A}\zeta_{+}\cD_{A}\phi+\zeta_{+}S\phi- \i F\,\zeta'_{+}+\i \frac{2}{3}q\phi\,\Gamma^{A}\cD_{A}\zeta_{+}\, , \nn\\
&&\delta_{+}\psi'_{+}=\i\Gamma^{A}\zeta'_{+}\cD_{A}\bar\phi+\zeta'_{+}S\bar\phi-\i \bar F\,\zeta_{+}+\i\frac{2}{3}q\bar\phi\,\Gamma^{A}\cD_{A}\zeta'_{+} \, , \\
&&\delta_{+}F=\i2 \wt\zeta_{+}\Gamma^{A}\cD_{A}\psi_{+}-2 S\,\wt\zeta_{+}\psi_{+}-\i2 \phi\,\wt\zeta_{+}\lambda_{+}+\i \frac{2}{3}(2q-1)\wt{\cD_{A}\zeta_{+}}\Gamma^{A}\psi_{+}\, , \nn\\
&&\delta_{+}\bar F=\i2 \wt\zeta'_{+}\Gamma^{A}\cD_{A}\psi'_{+}-2 S\,\wt\zeta'_{+}\psi'_{+} +\i 2\bar\phi\,\wt\zeta'_{+}\lambda'_{+} +\i \frac{2}{3}(2q-1)\wt{\cD_{A}\zeta'_{+}}\Gamma^{A}\psi'_{+} \, . \nn
\eea
This supersymmetry transformation rule is  consistent with the reality condition \eqref{realitySUSY} and
\be
\phi^\ast = \overline{\phi}\,,~~~F^\ast =\overline{F}\,, 
\ee
together with the 3-dimensional realisation of  the symplectic Majorana condition  \eqref{3dSM}. But again, as  we give up the condition  \eqref{3dSM}, we may use \eqref{realityPI} and 
\be
\phi^\ast = -\overline{\phi}\,,~~~F^\ast =-\overline{F}\,,~~~S^\ast =-S
\ee
to insure the action be positive definite. 
\\

The Lagrangian for a single chiral multiplet with R-charge $q$ is 
\bea\label{ChiralAction}
\mathcal L_{\text{chiral}}&=&\cD_{A}\bar\phi \cD_{A}\phi-S^{2}\bar\phi\phi+(2q-1)S\bar\phi\phi-q(q-2)\bar\phi\phi-(D_{3}+\p_{n}P)\bar\phi \phi-\bar FF\nn\\
&&-2\wt\psi'\Gamma^{A}\cD_{A}\psi-2\i S\wt\psi'\psi+\i(2q-1)\wt\psi'\psi+2\wt\psi'\lambda\phi-2\bar\phi\wt\lambda'\psi \ .
\eea
Here $q$ is the R-charge. The scalar field $\phi$ has R-charge $-q$, $\psi_{+}$ has R-charge $(-q+1)$. 
Boundedness of the potential for the scalar requires that $0 < q < 2$.  
Also here covariant derivatives are
\be
\cD_{A}\phi=\p_{A}\phi+\i A_{A}\phi,\quad \cD_{A}\bar\phi=\p_{A}\bar\phi-\i A_{A}\bar\phi,\quad \cD_{A}\psi=\nabla_{A}\psi+\i A_{A}\psi,\quad \cD_{A}\psi'=\nabla_{A}\psi'-\i A_{A}\psi'
\ee
In fact, the above action \eqref{ChiralAction} is Q-exact up to boundary terms
\be
\frac{1}{\wt\zeta_{+}\zeta'_{+}}\delta_{\zeta'_{+}}\delta_{\zeta_{+}}\Big[\wt\psi'_{+}\psi_{+}-\i(S\bar\phi\phi)+\i(q-1)(\bar\phi\phi)\Big]=\mathcal L_{\text{chiral}}+\text{boundary terms}
\ee

\section*{}

In order to perform the localization computation, we deform the partition function by a $Q$-exact term, $QV$ with a standard choice of the functional $V$ and a real parameter $t$. The path integral does not depend on the parameter $t$, and therefore, we can evaluate the partition function in the limit $t\rightarrow\infty$. In this limit, the partition function reduces to an integral over the solutions of the equations
\be
\delta\lambda^{i}=0,\quad \delta_{+}\psi_{+}=0\,,
\ee
which parameterize the localization background together with the measure that is given by the classical action evaluated on the localization background and the one loop determinant coming from the quadratic fluctuations of $QV$.  In the above, 
$\delta$ is the supercharge which reduces to $\delta_{+}$ at $r=\frac{\pi}{2}$. The solutions of the above equations are 
\be
\p_{A}S=0\,,\qquad \text{and}\qquad  S = D_3\,,
\ee
with all the other fields set to zero. Since we are dealing with an abelian gauge theory coupled to chirals at the boundary, at the quadratic order in fluctuations about the localization background, vector multiplet and chiral multiplet fluctuations decouple. 
The Lagrangian for the chiral multiplet \eqref{ChiralAction} which is itself $Q$-exact makes this decoupling manifest. 
The one loop calculation for the chiral multiplet is a free field calculation where the mass depends 
on the background value of the vector multiplet scalar $S$. 
Similar statements hold true for the fluctuations of the vector multiplet. The $QV$ functional of the vector multiplet is constructed in the standard way from $\delta(\lambda^{i\dagger}\delta\lambda^{i})$ \cite{Pestun:2007rz}. 
Since we are interested in an abelian vector multiplet, the fluctuations in $\delta(\lambda^{i\dagger}\delta\lambda^{i})$ are quadratic in the vector multiplet fields. 
The quadratic nature of the fluctuation determinant follows from the fact that
$\delta \lambda^{i}$ is itself linear in the fluctuations.

The issue of gauge fixing remains.  We follow again the lead of ref.\ \cite{Pestun:2007rz}. 
A convenient choice of supersymmetric gauge fixing Lagrangian is
\be
\mathcal L_{\rm ghost}=\hat\delta\Big(i\tilde c\nabla_{\mu} A^{\mu}+\xi\tilde c\,b\Big)\,.
\ee
where $\xi$ is a constant, and $\tilde c$ and $b$ are the BRST ghost fields.
The  SUSY variation $\delta$ needs to modified to $\hat \delta$ to include the BRST transformation.
At the end of the day, the vector multiplet one loop 
computation decouples from the physics on the boundary and 
does not give anything interesting except the anomaly, which we will discuss more below.

\section{The Localization Integral}
\label{sec:localization}

The localization result for our hemisphere partition function has a relatively simple form.
From setting $\delta_+ \lambda_+ = 0 = \delta_+ \psi_+$, we see that 
the classical piece of the action localizes to $\p_{A}S=0$ and $S = D_3$ with all the other fields set to zero.  There is thus a remaining
tree level contribution of the form
\be
I_{\rm tree} = \i \pi  \tau S^2 = \i \pi   \left( \frac{2 \pi i }{g^2} + \frac{\theta}{2\pi} \right) S^2 \ .
\ee
There is a one-loop contribution from the $n_+$ and $n_-$ chiral multiplets on the boundary, with R-charge $q_\pm$, which gives a measure factor to the
path integral \cite{Jafferis:2010un}:
\be
\exp \left[ n_+ \ell( 1 - q_+ + S) +  n_- \ell(1 - q_-  -S)  \right]\ ,
\ee
where\footnote{%
Note that for $S = \i \sigma$, $q_+ = q_- = \frac{1}{2}$ and $n_+ = n_- = n$, the measure factor reduces to
\be
\exp \left[  n \ell\left( \frac{1}{2} + \i \sigma\right) + n \ell\left( \frac{1}{2} -\i \sigma\right)\right] = 
\frac{1}{\left[ 2 \cosh (\pi \sigma)\right]^n} \ .
\ee
}
\be
\ell(z) = -z \log \left( 1 - e^{2 \pi \i z} \right) + \frac{\i}{2} \left( \pi z^2 + \frac{1}{\pi} \Li_2 \left( e^{2 \pi \i z} \right) \right) - \frac{ \i \pi}{12} \ .
\ee
In principle, the R-charges for each chiral multiplet could be taken independent, yielding $n_+ + n_-$ parameters in all.  
Indeed we will need to consider these extra degrees of freedom 
in calculating the partition function for the $n_+ = 2$, $n_- = 0$ theory below. 
For simplicity at this stage, we will use the permutation symmetry to introduce only two parameters $q_\pm$.  
The arbitrariness of $q_\pm$ reflects the fact that the true R-symmetry of the conformal theory is in general a mixture
of the naive R-symmetry and the U(1) symmetries carried by the chiral multiplets.

So far we have neglected the FI term (\ref{FI}).  While a nonzero real FI term is dimensionful and will in general spoil conformal invariance, a complexified FI term has an imaginary component associated with the mixing between the R-symmetry and the topological symmetry carried by monopole operators.  
The conserved current for the topological symmetry 
is the 3d Hodge dual of the boundary limit of the field strength.
These monopole 
operators contribute to the partition function a factor we parametrize as $e^{-2 \i  \pi q_t S}$ \cite{Jafferis:2010un}
(see also \cite{Pufu:2016zxm} near (3.13)). 

The contour of integration is to take $S$ to be pure imaginary, from $-\i \infty$ to $\i \infty$. 
Defining $S = \i \sigma$, our partition function is then\footnote{%
 For the Euclidean action, the path integral is defined schematically via $Z \equiv \int [d \phi] e^{-I[\phi]}$. 
}
\be
Z = \int \d \sigma \, e^{\i \pi \tau \sigma^2} \exp \left[ n_+  \ell( 1 - q_+ + \i \sigma) + n_- \ell(1 - q_-  -\i \sigma) + 2 \pi q_t \sigma \right] \ .
\label{Z}
\ee
This form of the partition function was conjectured in \cite{Gaiotto:2014gha}.
In the limit where $\tau$ is purely real and equal to an integer $k$ (and $n_+ - n_-$ is even), we have recovered the 
localization result for the $S^3$ partition function of an ${\mathcal N}=2$ supersymmetric, level $k$, abelian Chern-Simons theory \cite{Kapustin:2009kz,Jafferis:2010un}.  (In the case $n_+ - n_-$ is odd, then $k$ should be equal to an integer plus one half to cancel an anomaly.)
The novelty here is that we can make sense not just of integer and half integer $\tau$ but of any value of 
$\tau$ in the upper half plane.  The integer and half integer 
values of $\tau$ correspond to ``decoupling'' limits where we can
recover a pure 3d interpretation of this 4d system.

We have neglected a one-loop contribution from the 
vector multiplet in the bulk.  
We shall largely ignore it after a brief discussion here.
This one-loop contribution is logarithmically divergent and gives rise to a scale anomaly, 
$\log Z \sim a \log (R / \epsilon)$ where $R$ is the radius of the hemisphere, $\epsilon$ is a short distance cut-off, and $a$ is the $a$-anomaly coefficient associated with the ${\mathcal N}=2$ vector multiplet.
This $a$ is precisely the anomaly coefficient proven to be monotonic under renormalization group flow
\cite{Komargodski:2011vj}.   
An issue for us is that this divergence may hide some $\tau$ and $\bar \tau$ dependence.
A sensible way to regulate this divergence is to divide by the partition function on the full sphere \cite{Gaiotto:2014gha}:
\be
\label{Fpartial}
2 F_\partial \equiv - \log \frac{|Z|^2}{Z_{S^4}} \ .
\ee
The ratio is manifestly finite and can be used to identify coupling dependence hiding in the choice of $\epsilon$. 
For example, in the weakly coupled (large $\tau$) case, $Z \sim \tau^{-1/2}$ and therefore $\bar Z \sim \bar \tau^{-1/2}$. 
The partition function for a free Maxwell field on the sphere is known to scale as $Z_{S^4} \sim (\tau - \bar \tau)^{-1/2}$, leading to
\be
e^{-F_{\partial}} \sim \frac{| \tau - \bar \tau|^{1/4}}{|\tau |^{1/2}} \ .
\ee
This regulated partition function has the correct $\tau$ and $\bar \tau$ dependence and gives rise to the correct integrated one point functions  $\langle F_{\mu\nu} F^{\mu\nu} \rangle$ and $\langle F_{\mu\nu} \tilde F^{\mu\nu} \rangle$ \cite{DiPietro:2019hqe}.\footnote{%
We have not included the contributions of point like instantons in \eqref{Z}. These contributions come from the singular field configurations of the gauge field localized at the north pole \cite{Pestun:2007rz} and also depend on the complexified parameter $\tau$. However, point instanton contributions cancel out in the definition of the regulated free energy \eqref{Fpartial}. 
}
Despite this subtlety, we shall proceed by studying $Z$ on its own, cognizant that its dependence on $\tau$ and lack of dependence on $\bar \tau$ may be misleading in certain cases.

To determine the $q_\pm$ and $q_t$, it is well known that in the purely 3d case, 
one must minimize $|Z|$ as a function of the potentials $q_i$
\cite{Jafferis:2010un,Closset:2012vg}.  
Because the 4d part our theory is a free Maxwell field and its super partners, we expect 
the minimization arguments generalize
to our case.\footnote{%
 Ref.\ \cite{Gaiotto:2014gha} argues for this minimization principle by mocking up the effects of the 4d gauge field
 in a purely 3d example by coupling the matter sector of interest to a large number of weakly charged chiral multiplets.
}
Something that will prove very useful for us is that
the second derivatives of $Z$ with respect to the $q_i$ determine current two-point functions of the 
associated symmetries that mix with the R-symmetry.  Minimization morally means that the one-point functions
of the currents vanish and the two-point functions are reflection positive.  

In flat  space, the two point function of a 
current is given by
\be
\langle j_{\mu}(x)j_{\nu}(0) \rangle=\frac{\Pi}{16\pi^{2}}(\delta_{\mu\nu}\p^{2}-\p_{\mu}\p_{\nu})\frac{1}{x^{2}}+\frac{i\kappa}{2\pi}\epsilon_{\mu\nu\rho}\p^{\rho}\delta(x)\, ,
\ee
where $\Pi$ and $\kappa$ can be determined by the $S^{3}$ partition function of the CFT as \cite{Closset:2012vg}
\be
\Pi=\frac{1}{\pi^{2}|Z|^{2}}\p^{2}_{q}|Z|^{2}\Big|_{q=q_{*}},\quad \kappa= -\frac{1}{2\pi }\text{Im}\Big[\frac{1}{Z}\p_{q}^{2}Z\Big]\Big|_{q=q_{*}}\,.
\label{Pikappadef}
\ee
or equivalently
\be
\Sigma \equiv \kappa + \frac{\i \pi}{4} \Pi = \left. \frac{\i }{2\pi} \partial_q^2 \log Z \right|_{q = q_*} \ .
\ee
Because the 4d part of our theory is free, we expect 
these results generalize to our $HS^4$ case.
Here $q_{*}$ is the R-charge extremizing the partition function. Note $\Pi$ is normalized such that each chiral superfield contributes one in the weak coupling limit (as can be verified for example from the photon self energy \cite{Herzog:2018lqz} in these theories).  These relations are straighforwardly generalized to multiple $q_i$, $\Sigma_{ij} = \frac{\i}{2\pi} \partial_i \partial_j \log Z$, evaluated at the critical point now for both $q_i$ and $q_j$. 

 Through Kubo formulae, the constants $\Pi$ and $\kappa$ additionally 
 determine the conductivity with respect to an  electric field that couples
 to the current.  In particular, $\Pi = 8 \sigma$ and $\kappa =  2 \pi \sigma_H$ 
 where $\sigma$ and $\sigma_H$ are the regular and Hall conductivity respectively, $\Sigma = 2\pi (\sigma_H + \i \sigma)$.
  Such conductivities are discussed at length in a condensed matter context where they are relevant for describing systems at a quantum phase transition, for example thin films at a superconductor-insulator transition \cite{Fisher:1990zza,Wen:1990gi}.\footnote{%
 The field strength is often normalized so that the coupling $g$ appears not in the kinetic term of the action but
 with the interaction term.  This redefinition will introduce a factor $g^2$ into the conductivity, $\sigma \to g^2 \sigma $ and $\sigma_H \to g^2 \sigma_H$.  A true physicist might also want to reintroduce $\hbar$, which we have set equal to one, rescaling the conductivity now not by $g^2$ but by $g^2 / \hbar$.  
  }

Let us make the following change of variables $q_+ + q_- = 2 q_f$ and $q_+ - q_- = 2 q_g$, $f$ for flavor and $g$ for gauge.  We identify the difference $q_+ - q_-$ with the gauge symmetry because the $+$ chiral multiplet has charge $+q_g$ while the $-$ chiral has charge $-q_g$.
We can absorb the $q_g$ inside the $\ell(z)$ functions by making the change of variables $\sigma \to \sigma - i q_g$.  
In the presence of a nonzero $\tau$, this shift leads to the following extra term in the exponent of the integrand
\be
2 \pi \tau q_g \sigma - \i \pi \tau q_g^2 -2 \i \pi q_t q_g  \ .
\ee
If we think of our partition function as a function of two variables $Z(q_g, q_t)$ and we assume 
that $q_g$ is small enough that no poles are crossed by  shifting the contour, the shift in the exponent
induces in the following relation, 
\be
\label{ZZequality}
Z(q_g,q_t) = Z(0, q_t + \tau q_g) e^{-\i \pi \tau q_g^2 - 2 \i \pi q_t q_g} \ .
\ee
The reality properties of $\tau$ play a significant role here.  In the decoupling limit where $\tau$ is a real (and nonzero) parameter, this equality implies the partition function depends on the single real quantity $q_t + \tau q_g$, up to some complex phases.  For $\tau$ complex, on the other hand, there are two real degrees of freedom in $q_t + \tau q_g$.

This relation (\ref{ZZequality}) leads then to the following 
identification between second derivatives, 
\be
\label{sumrule}
\Sigma_{gg} = \tau + \tau^2 \Sigma_{tt} \ , \; \; \; \Sigma_{gg} = \tau \Sigma_{gt} \ .
\ee
A similar result can be found in \cite{Gaiotto:2014gha} as a sum rule relating a two-point function of the gauge current to a two-point function of the topological current.  Their derivation was very different, relying on conformal symmetry and
a crossing constraint on the two point function for the field strength.

There is a second intriguing identity that follows from the structure of the partition function, namely
\be
\label{idtwo}
\partial_{q_t}^2 Z = - 4 \pi \i \partial_\tau Z \ .
\ee
Physically, this equality should relate integrated one point functions $\langle F_{\mu\nu} F^{\mu\nu} \rangle$ and
$\langle F_{\mu\nu} \tilde F^{\mu\nu} \rangle$ to the two-point function of the topological current.  Such a relation was also 
given in \cite{Gaiotto:2014gha}.  Unfortunately, our relation does not match theirs on the nose because of the damage we did in neglecting the one-loop contribution of the vector multiplet.  As a statement about (\ref{Z}), however, (\ref{idtwo}) is correct.

We begin with an extensive set of saddle point analyses of $Z$, and follow with a contour
integral method that can be used to evaluate $Z$ in the decoupling limits, where $\tau$ is a rational number.
We save a lengthier discussion of duality and transport for section \ref{sec:sduality}.

\subsection{Saddle Point Analysis}

We restrict to the simple case $n_+ = n_- = n$, in the limits of large $|\tau|$ and large $n$. 
There is a saddle point at $\sigma = 0$ for any value of $\tau$ or $n$.  
We can in principle minimize $|Z|^2$ as a function of both $q_g$ and $q_f$.  
Under a symmetry exchanging the $+$ and $-$ chiral fields, 
there must be an extremum at $q_g = 0$ although it may not in general be a minimum.
In the limits of large $|\tau|$ and large $n$, there are critical points for $|Z|^2$ 
near $q_f = 1/2$ and 3/2. 
Only the critical point near $q_f = 1/2$ turns out to be a local minimum. The critical point near $q_f=3/2$ is a maximum at large $|\tau|$ and a saddle at large $n$.
 We did not perform an exhaustive search for critical points in $(q_g, q_t)$ space.  Note however that for stability of the theory, $0 < q_\pm < 2$.

\begin{figure}
\begin{center}
\includegraphics[width=2.5in]{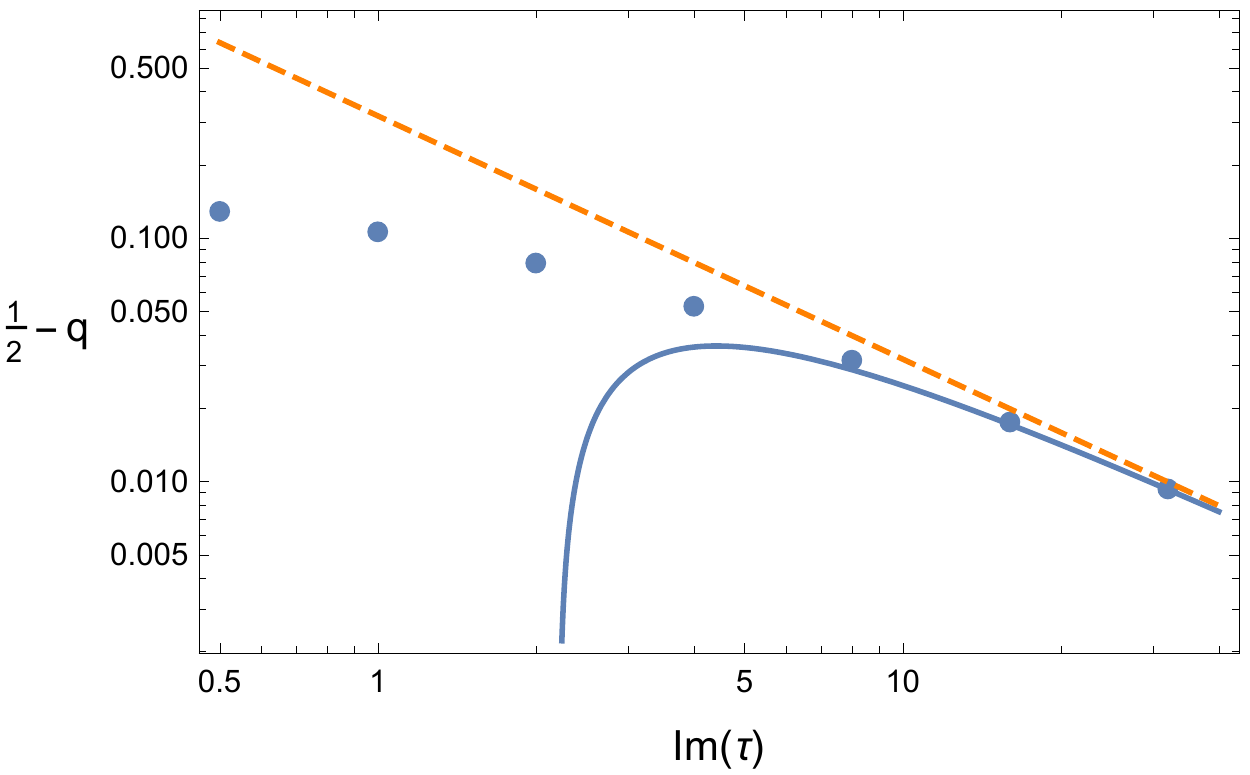}
\includegraphics[width=2.5in]{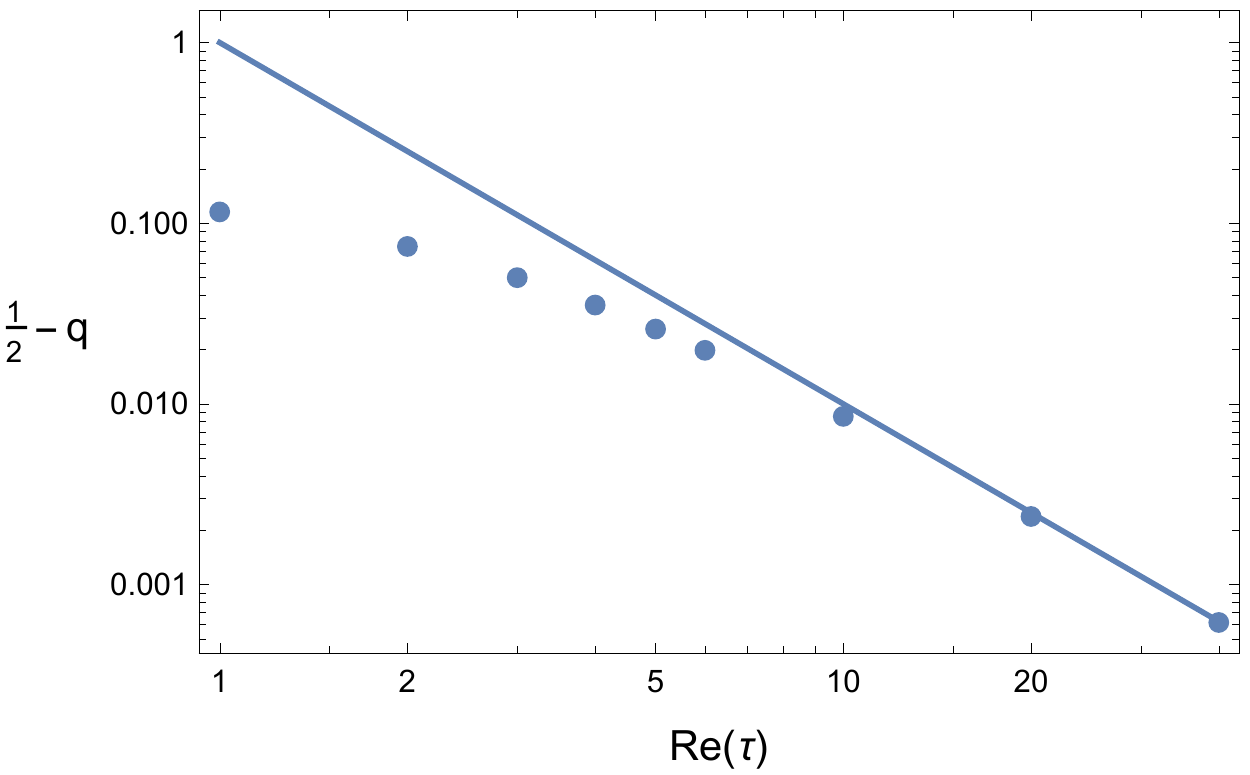}
\end{center}
\caption{
The points are numerically determined values for $q$ at different values of $\tau$ and with $n=1$.  
The straight lines are the nonzero leading order saddle point approximations
for $\frac{1}{2}-q$ while the blue curve on the left is the second order correction.
In the plot on the left $\operatorname{Re}\tau = 0$ while in the plot on the right
$\operatorname{Im} \tau = 0$.  
\label{largetauplot}
}
\end{figure}

\begin{figure}
\begin{center}
\includegraphics[width=3in]{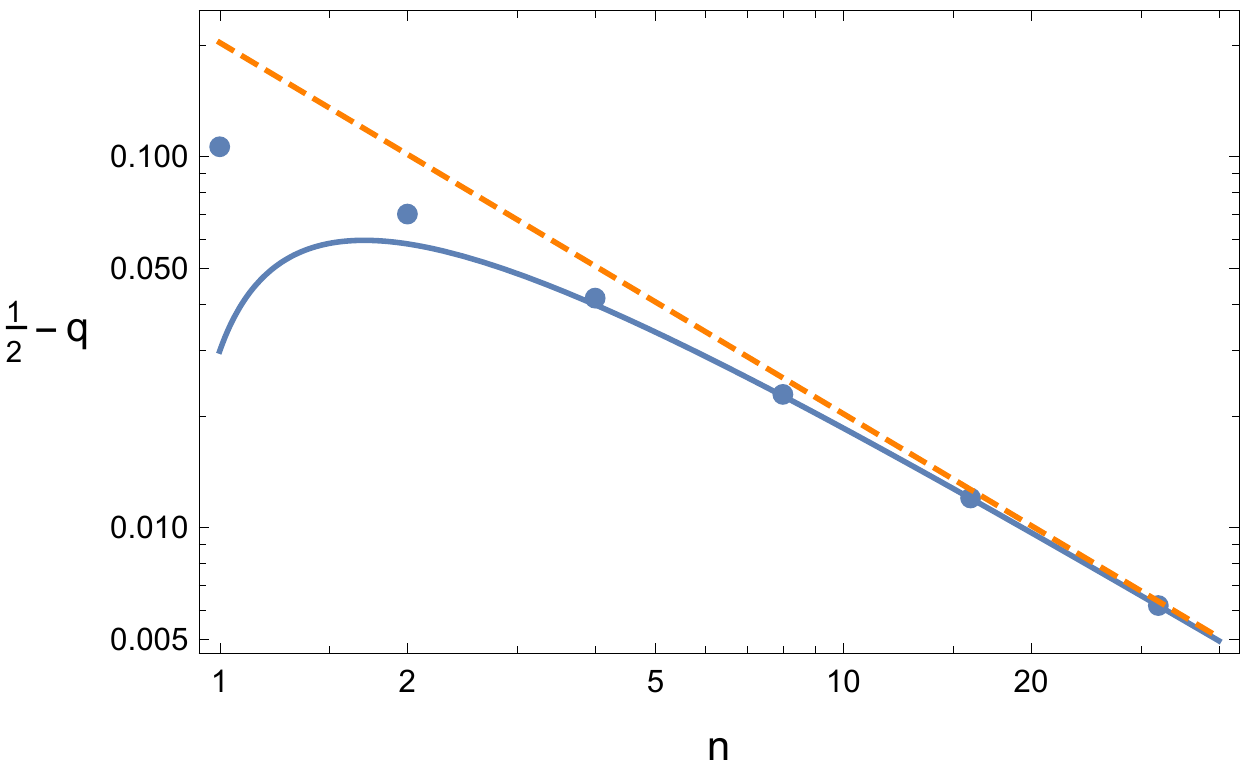}
\end{center}
\caption{
\label{largenplot}
The points are numerically determined values for $q$ at various values of $n$ for $\tau = i$.  
The straight and curved lines are the first and second order saddle point approximations 
respectively for
$\frac{1}{2}-q$ at large $n$.  
}
\end{figure}

\begin{figure}
\begin{center}
\includegraphics[width=2.5in]{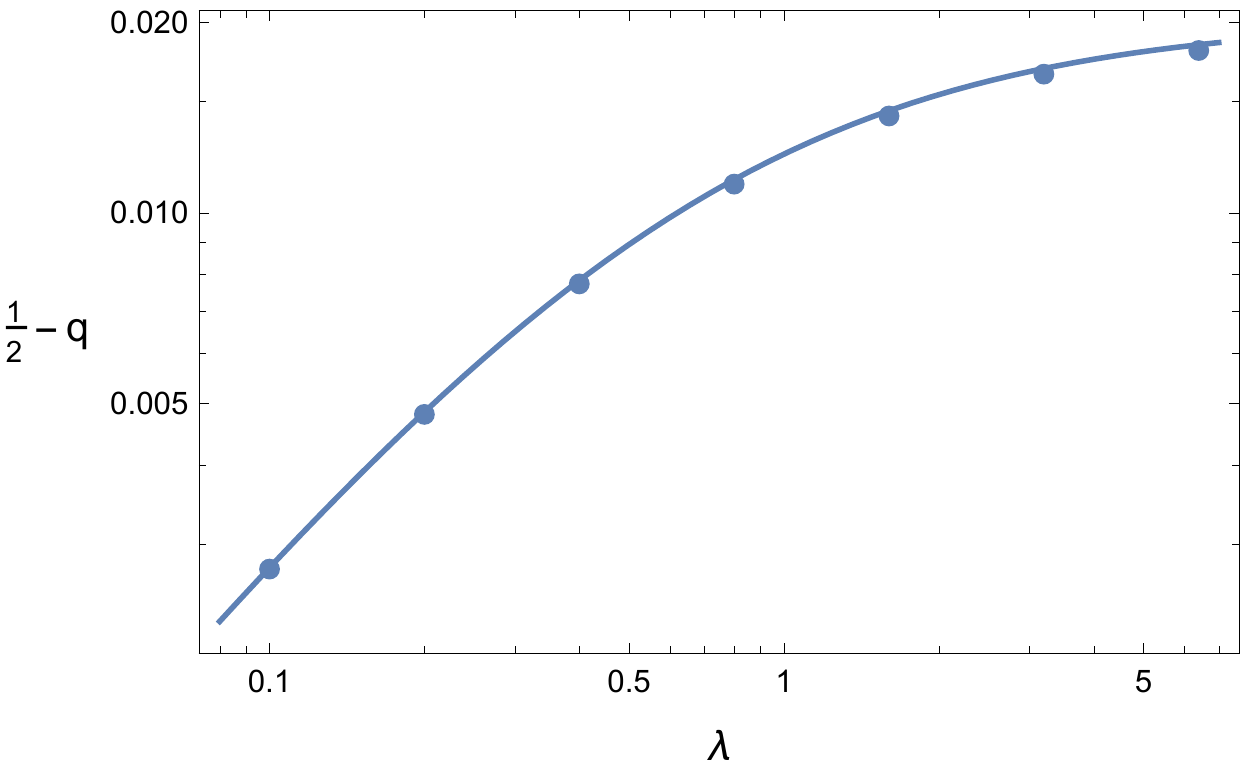}
\includegraphics[width=2.5in]{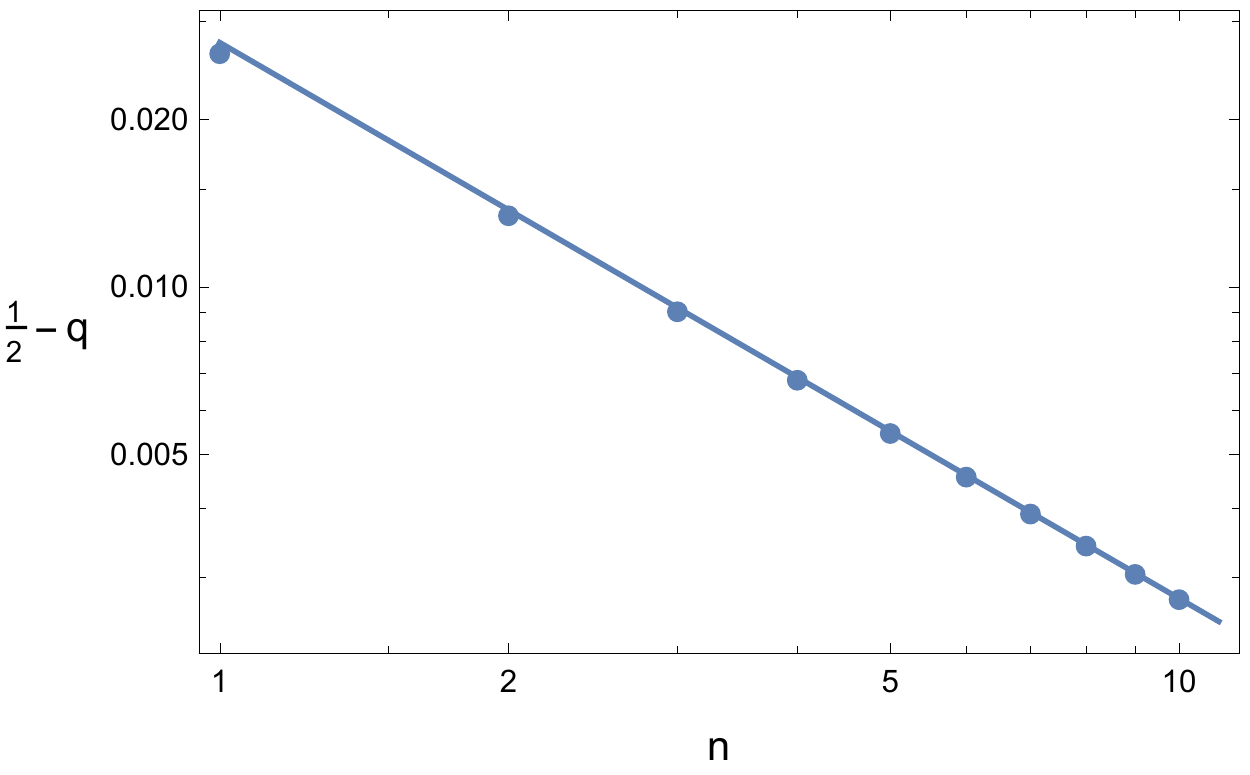}
\end{center}
\caption{The points are numerically determined values for $q$ and the straight line is the first order saddle point approximation in the fixed $\lambda$, large $n$ limit for $\alpha = \frac{\pi}{2}$:
(left)  various values of $\lambda$ and $n=10$; 
(right) various values of $n$ and $\lambda = 0.1$.
\label{lambdaplot}}
\end{figure}

\subsubsection*{Large $\tau$}

When $|\tau|$ is large, parametrizing $\tau = |\tau| e^{i \alpha}$, expanding to quartic order around the saddle point, we find $|Z|$ is minimized when
\be
q_f = \frac{1}{2} - \frac{\sin \alpha}{\pi |\tau|} - \frac{ \pi^2 - 4 +(4+ (1+2n) \pi^2) \cos(2 \alpha)}{4 \pi^2 |\tau|^2} + O(|\tau|^{-3}) \ .
\ee
The decoupled case $\alpha = 0$ matches the large Chern-Simons level 
calculation performed at the end of \cite{Jafferis:2010un}.
We have confirmed this formula for certain special cases ($\alpha = 0$ and $\alpha = \pi/2$ with $n=1$) through
numerical integration of (\ref{Z}) (see figure \ref{largetauplot}).
At the critical point
\be
|Z|^2 = \frac{1}{2^{2n} |\tau|} - \frac{\pi n \sin \alpha}{2^{2n+1} |\tau|^2} + \frac{n(\pi^2 n-8 + (8 - (2+3n) \pi^2) \cos (2 \alpha))}{2^{2n+4} |\tau|^3} + O(|\tau|^{-4}) \ .
\ee

We can furthermore compute the two-point functions of the gauge and flavor currents.  We leave it to the reader to
apply the sum rule (\ref{sumrule}) and compute the two-point function of the topological current:
 \bea
\label{largetauPi}
&&\Pi_{ff} =2n-\frac{n(\pi^{2}-8)\sin\alpha}{\pi|\tau|}
+ \frac{-2n(8 + \pi^2) + n (16 + 10 \pi^2 - (2+n) \pi^4) \cos(2 \alpha)}{2 \pi^2 |\tau|^2}
+\mathcal O(|\tau|^{-3})\,,\nn\\
&&\kappa_{ff} =\frac{n\pi^{2}\cos\alpha}{4|\tau|}-\frac{n\pi\Big(2\pi^{2}-8+8n+n\pi^{2}\Big)\sin2\alpha}{8|\tau|^{2}}+\mathcal O(|\tau|^{-3})\, , \\
&&\Pi_{gg} = 2n - \frac{n((n+1) \pi^2-8 ) \sin(\alpha)}{\pi |\tau|} + O(|\tau|^{-2}) \ , \nonumber
\\
&& \kappa_{gg} =
\frac{\pi^2 n ( 8n + (n+2) \pi^2) \cos(\alpha)}{|\tau|} 
+ O(|\tau|^{-2}) \ .
\label{Piglargetau}
\eea
Note that $\Pi_{ff}$ and $\Pi_{gg}$, which are proportional to conductivities, 
decrease as the coupling $1/\tau$ increases.
This behavior is in accord with physical intuition, that increased scattering between quasi-particles in a weakly coupled
description 
will hinder transport.  
The Hall conductivity which is proportional to 
$\kappa$, on the other hand, has a sign determined by the real part of $\tau$.
These results are in principle accessible via perturbation theory.  In practice, even the leading order term involves a sum over several one loop diagrams.

\subsubsection*{Large $n$}

We can equivalently perform a saddle point integration in the limit of large $n$.  In this case, we find instead
\be
q_f = \frac{1}{2} 
- \frac{2}{\pi^2 n} 
+ \frac{2 \left(12 -\pi^2 +2 \pi \operatorname{Im} \tau \right)}{\pi^4 n^2}
 + O(n^{-3}) \ .
\ee
We confirmed this result numerically, for the special case $\tau = i$ (see figure \ref{largenplot}).  
At the critical point,
\be
|Z|^2 = 2^{-2n} \left(  \frac{2}{\pi n} - \frac{ 8 - \pi^2 + 4 \pi \operatorname{Im} \tau}{\pi^3 n^2} + O(n^{-3}) \right) \ .
\ee
The associated current two-point functions are
\bea
\label{largenPi}
\Pi_{ff}&=& 2n+\frac{2}{\pi^{2}}(16-\pi^{2})+
\frac{2(-224 + 12 \pi^2 + \pi^4 - 2 \pi(24 -\pi^2)) |\tau| \sin(\alpha)}{\pi^4 n} +
\mathcal O(n^{-2})\,,
\nn\\
\kappa_{ff}&=&-\frac{(16-\pi^{2})|\tau|\cos\alpha}{\pi^{2}n}
+\mathcal O(n^{-2})\,, \\
\Sigma_{gg} &=& \tau + \frac{2 \i \tau^2}{2 n} +\frac{2 (\pi^2 - 8) \i \tau^2 - 4 \pi \tau^3}{\pi^3 n^2} + O(n^{-3}) \ .
\label{Piglargen}
\end{eqnarray}
Applying the sum rule (\ref{sumrule}) to these results, we find that
 $\Sigma_{tt}$ is $O(n^{-1})$.

\subsubsection*{Large $\tau$ and large $n$}

In the case when $n$ and $|\tau|$ are both large, we can consider a perturbative expansion in $\frac{1}{n}$ keeping the ratio $\frac{n}{|\tau|}=\lambda$ fixed. In this case, the partition function extremizes for
\be
q_f=\frac{1}{2}-\frac{2\lambda(\pi\lambda+2\sin\alpha)}{\pi(4+\pi^{2}\lambda^{2}+4\pi\lambda\sin\alpha)n}+\mathcal O(n^{-2})\,,
\ee
where $\tau=|\tau|e^{i\alpha}$. 
The agreement of this result with numerical evaluation of the partition function is remarkably good; 
see figure \ref{lambdaplot}.
The partition function for the above value of the R-charge and $\alpha=\frac{\pi}{2}$ is given by
\be
|Z|^{2}=\frac{\lambda}{4^{n}n(2+\pi\lambda)}\Big[2+\frac{(-8+\pi^{2})\lambda^{2}}{n(2+\pi\lambda)^{2}}+\mathcal O(n^{-2})\Big]\,.
\ee
The flavor current two-point function in this limit is
\bea
\Pi_{ff}&=&2n-\frac{2\lambda}{\pi(4+\pi^{2}\lambda^{2}+4\pi\lambda\sin\alpha)^{2}}\Big(\pi\lambda(-64+\pi^{4}\lambda^{2}+\pi^{2}(8-16\lambda^{2})) \nn \\
&&+4\pi(16-\pi^{2})\lambda\cos2\alpha+(-64+6\pi^{4}\lambda^{2}+8\pi^{2}(1-10\lambda^{2}))\sin\alpha\Big)+\mathcal O(n^{-1})\,. \nn \\
\label{Pidouble}
\kappa_{ff} &=& \frac{\pi\lambda\cos\alpha\Big[\pi(4-(16-\pi^{2})\lambda^{2})-4(8-\pi^{2})\lambda\sin\alpha\Big]}{(4+\pi^{2}\lambda^{2}+4\pi\lambda\sin\alpha)^{2}}+\mathcal O(n^{-1})\,.
\eea
\ndt In particular, for $\alpha=\frac{\pi}{2}$, $\kappa$ must vanish by parity, and the above expression for $\Pi$ 
simplifies to 
\be
\Pi_{ff}=2n+\frac{2\lambda\Big(16-2\pi^{2}+\pi(16-\pi^{2})\lambda\Big)}{\pi(2+\pi\lambda)^{2}}+\mathcal O(n^{-1})\,,
\ee

\begin{figure}
\begin{center}
\includegraphics[width=3in]{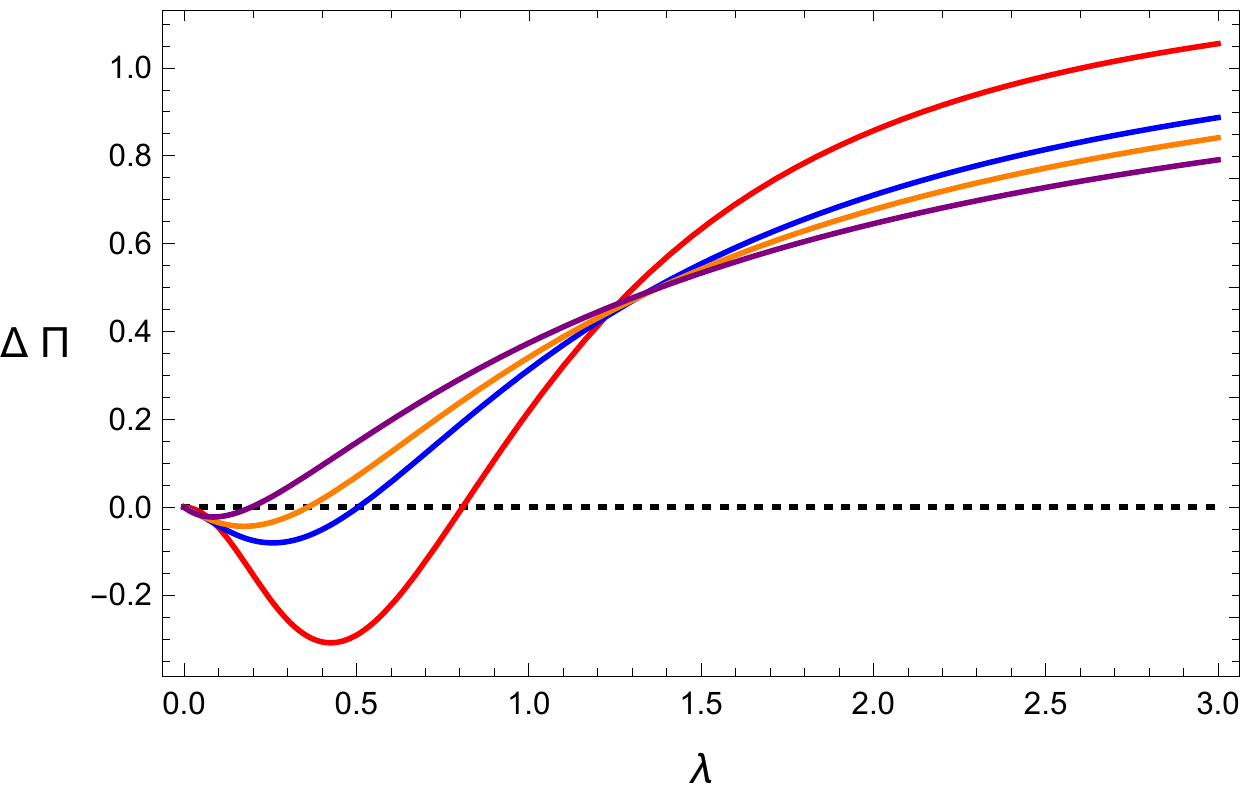}
\end{center}
\caption{A plot of $\Delta \Pi = \Pi_{ff} - 2n$ in the large $n$ and $\tau$ limit with $\lambda = n / |\tau|$ fixed.  From top to bottom on the right, the curves correspond to $\alpha = 0$, $\frac{\pi}{6}$, $\frac{\pi}{4}$, and $\frac{\pi}{2}$.  
The minimum at $\alpha = 0$ occurs when $\lambda = \sqrt{\frac{2}{32-\pi^2}}$ and $\Delta \Pi = - \frac{\pi^2}{32}$.  
The minimum at $\alpha = \frac{\pi}{2}$ occurs at $(\lambda, \Delta \Pi) = \left( \frac{2\pi^2-16}{\pi(24 - \pi^2)}, - \frac{(\pi^2-8)^2}{16 \pi^2} \right)$.
\label{fig:deltaPiplot}}
\end{figure}

The behavior of $\Pi_{ff}$ in the double scaling limit (\ref{Pidouble}) is interesting; see figure \ref{fig:deltaPiplot}.  The result (\ref{Pidouble}) reproduces (\ref{largetauPi}) when $\lambda \to 0$ and (\ref{largenPi}) when $\lambda \to \infty$.  For small $\lambda$, increasing the coupling $\lambda$ acts to decrease the transport coefficient $\Pi_{ff}$ as expected.  However, as $\tau$ gets smaller, there is eventually an inversion and $\Pi_{ff}$ starts to get larger as the coupling increases.  
This type of behavior suggests that as $\lambda$ increases, there may at some point be
a better weakly coupled description with a dual $\tau'$.  Increasing $|\tau'|$ would correspond to decreasing $|\tau|$.
Thus $\Pi$ again follows our intuition but with respect to the weakly coupled description.  
We shall return to this idea of a dual description in the next section.\footnote{%
  For related calculations of conductivities in nonabelian and non-supersymmetric Chern-Simons theories, 
  see \cite{Gur-Ari:2016xff,Aharony:2012nh,GurAri:2012is}.  These references explore a large $N$ limit and
  the consequences of duality on transport.
 }

The result for the gauge conductivity interpolates between the results (\ref{Piglargetau}) at small $\lambda$ and 
(\ref{Piglargen}) and large $\lambda$: 
\begin{eqnarray}
\Pi_{gg} &=& \frac{4 n(2 + \pi \lambda \sin(\alpha)) }{4 + \pi^2 \lambda^2 + 4 \pi \lambda \sin(\alpha)} + O(n^0) \ , 
\\
\kappa_{gg} &=& \frac{ \pi^2 n \lambda \cos(\alpha)}{4 + \pi^2 \lambda^2 + 4 \pi \lambda \sin(\alpha)} + O(n^0) \ .
\end{eqnarray}
Already the first order term is nontrivial.

\subsection{Residue Method}

For rational values of $\tau$, the integral (\ref{Z}) can be performed by contour integration.
The central idea is to take advantage of a quasiperiodic property of the integrand for these special values.
(Indeed, this same method can be employed to perform the usual Gaussian integral by contour integration \cite{Kneser,Remmert}.)  Assume we have a definite integral
\be
\int f(\sigma) \d \sigma
\ee
and an extension of the integrand to the complex plane
with the quasi-periodic property that $f(z +a) f(z-a) = f(z)^2$.  
Assume furthermore that $f(z)$ dies off suitably fast as $|{\rm Re}(z)| \to \infty$. 
Then one can replace the original integral with a contour integral\footnote{%
 We would like to thank M.~Marino for introducing us to this contour method.
}
\be
\int f(\sigma) \d \sigma = \oint_C \frac{f(z)}{1- \frac{f(z)}{f(z-a)}} \d z
\ee
where the contour runs from $-\infty$ to $\infty$ on the real axis and then back from $+\infty$ to $-\infty$ along the line
${\rm Im}(z) = {\rm Im}(a)$.  The contour integral can be replaced in turn by a sum over poles inside the strip $0 < {\rm Im}(z) < {\rm Im}(a)$.  

In our case, 
\be
f(z) = e^{\i \pi \tau z^2 + 2 \pi q_t z } \exp \left[ n_+ \ell( 1 - q_+ + \i z) + n_- \ell(1 - q_- -\i z) \right] 
\ee
is quasiperiodic with respect to $a = i s$ when $s^2 \tau = r$ for $r$ and $s$ integers if $n_+ + n_-$
is even.  Otherwise, for $n_+ + n_-$ odd, we have the restriction $s^2 \left(\tau +\frac{1}{2}\right)= r$.  
Interestingly, for $s=1$, these cases correspond precisely to the decoupling limits, where the partition function
is equal to that of a purely 3d Chern-Simons theory, with the appropriate Chern-Simons level, either integer or half-integer
depending on the parity of $n_+ - n_-$.  

In general, the poles can be determined by solving for the roots of a polynomial.  
In the case $q_+ = q_- = 1/2$, the poles all lie on the imaginary axis and are easy to characterize. (See Appendix \ref{app:specialcases} for results pertaining to these cases.)  For 
more general values of the R-charges, 
the poles move off, and the story gets rich, especially, since we are supposed to extremize
with respect to $q_+$ and $q_-$.
Our strategy will be to limit
complexity by examining cases where finding the poles involves solving only a linear or quadratic equation.
We will consider 1) $n_+ = 1$ and $n_- = 0$; 2) $n_+ = 2$ and $n_-=0$; 3) $n_+ = 1$ and $n_-=1$; 4) $n_+=2$ and $n_- =2$. To simplify notation, we will present the results in terms of a single $q$ which is either $q_f$ in the cases $n_+ = n_-$ or $q_+$ when $n_- = 0$.  We will largely ignore the dependence on $q_g$ and $q_t$.  Note that as $\tau$ is real in this analysis, there is no distinction between $q_g$ and $q_t$ unles $\tau =0$.

The integral (\ref{Z}) appears also in studies of  hyperbolic geometry and integrable systems.  
In ref.\  \cite{Garoufalidis:2014ifa},  using this contour integration method, the authors provide
a residue formula for a generalization of (\ref{Z}).  In our context, 
the interpretation would be a
 partition function of an abelian Chern-Simons theory on a squashed three sphere, for rational
 value of the squashing parameter.  The sum over poles was then later cast as a sum over ``Bethe roots'', thus
 making a connection with integrability \cite{Closset:2018ghr}.  

\subsubsection*{$n_+ = 1$ and $n_- = 0$}

To compute the partition function by our residue method, 
the poles correspond to roots of the following algebraic equation in $x = e^{2 \pi \sigma}$:
\be
\label{firstpoly}
x^{1+r}  + x e^{\i \pi (r+q)} - e^{\i \pi (r-q)} = 0 \ ,
\ee
where we have set $\tau = r + \frac{1}{2}$.

This polynomial is already quite interesting from the point of view of hyperbolic geometry. The work \cite{Dimofte:2011ju} proposed an interesting correspondence between supersymmetric
gauge theories in three dimensions and the geometry of three manifolds.  One example that they study
closely, which forms in fact a building block to construct more complicated geometries and gauge theories,
 is the $n_+ = 1$ and $n_-=0$ theory with $\tau = \frac{1}{2}$.

We may consider the volumes of the three smallest hyperbolic three-manifolds.  They can be computed in the following way.  The volume is the imaginary part of an expression involving a dilogarithm:
\be
\operatorname{Im} \left( \Li_2(x) + \log |x| \log(1-x) \right)\ ,
\ee
and  let $x$ be the root of (\ref{firstpoly}) with a positive imaginary part.  In particular, for the smallest manifold, 
also called the Weeks manifold, 
we take $r=2$ and $q=1$ for which we need a root of $x^3 -x+1$.  The volume is $0.942707$.  
For the second smallest manifold, sometimes called the Thurston manifold, 
$r=3$ and $q=1$, for which we need a root of $x^4 + x - 1$.  The volume is $0.981369$.
Finally, the third smallest manifold comes from $r=1$ and $q=0$ with polynomial $x^2-x+1$.  Now the roots
are cube roots of $-1$, and the volume is $1.01494$.  

In fact, ref.\ \cite{Gang:2017lsr} already pointed out this relation.
They considered the partition function of this $n_+ = 1$ and $n_-=0$ theory
on a warped $S^3$, in the limit where the $S^3$ approaches $S^2 \times S^1$.
They noticed that the saddle point evaluation of the log of the partition function
 gave precisely these three minimal volumes.

To keep our life simple we will consider only $\tau = \frac{1}{2}$ and $\tau = \frac{3}{2}$ 
so that we do not need to solve more than a quadratic equation.
In the case $\tau = \frac{1}{2}$, the one pole is at
\be
x_0 = \frac{e^{-i \pi q}}{1 + e^{i \pi q}} \ .
\ee
The branch cut of the log can be adjusted to put the imaginary part of  $\sigma_0 \equiv \frac{1}{2\pi} \log x_0$ between zero and one, leading to the 
residue formula for the partition function $Z[\tau, q]$:
\be
\label{Zhalf}
Z\left[\frac{1}{2}, q \right] = \frac{i  e^{\frac{\i}{2} \pi \sigma_0^2 } }{1 + e^{\i \pi q}}\exp \left[ \ell(1-q+i \sigma_0) \right] \ .
\ee 
We can use some dilogarithm identities to re-express this result in a simpler way:
\be
\label{Zhalftwo}
Z\left[\frac{1}{2}, q \right] = \exp \left[ \ell \left( \frac{q+1}{2} \right)  - \frac{\i \pi}{24} (3q-1)^2 + \frac{\i \pi}{12} \right] \ .
\ee
We have recovered up to a phase the partition function of a single chiral field with R-charge $\frac{1-q}{2}$.
The only critical points of this partition function in the allowed range $0<q<2$ are at $q = 0$ and $q=2$.
As $q=2$ leads to a chiral superfield with dimension below the unitarity bound, we take $q=0$ and find a free chiral
with R-charge $1/2$.  This result is in agreement with \cite{Intriligator:2013lca}.

In the case $\tau = \frac{3}{2}$, we have two poles at
\be
x_{\pm} = \frac{1}{2} \left( e^{\i \pi  q} \pm e^{-\frac{\i}{2} \pi q}\sqrt{e^{3 \i \pi q}-4} \right)\ .
\ee
Now the partition function becomes a sum over two terms
\begin{eqnarray}
Z\left[ \frac{3}{2}, q \right] &=&  \frac{1}{4\pi} \left(  
\left( 1 + \frac{e^{\i \pi q/2}}{\sqrt{e^{3 \i \pi q}-4}} \right) e^{\frac{3}{2} \i \pi \sigma_+^2 +\ell(1-q+i \sigma_+) } +
\right.
\nonumber \\
&&
\left. 
\hspace{1in} + 
\left( 1 - \frac{e^{\i \pi q/2}}{\sqrt{e^{3 \i \pi q}-4}} \right) e^{\frac{3}{2} \i \pi \sigma_-^2 +\ell(1-q+i \sigma_-)} 
\right)
\end{eqnarray}
where $\sigma_\pm \equiv \frac{1}{2\pi} \log x_\pm$.  
Here the absolute value of the partition function is minimized at $q=\frac{1}{3}$.  There is a local minimum as well, at $q = 1.95185$.  We do not have a good understanding why there are two minima in this case.

\subsubsection*{$n_+ = 2$ and $n_- = 0$}

As $n_+ + n_-$ is even, we take instead $\tau=r$ to be an integer. 
The algebraic equation determining the location of the poles is
\be
x^{1+r} e^{2 \i \pi q} +x^2 e^{\i \pi(4 q + r)} - 2 x e^{\i \pi(2q+r)} + e^{\i \pi r} = 0
\ee
Already when $\tau =0$, there are two poles:
\be
x_\pm= \frac{1}{2} (1 \pm i \sqrt{3}) e^{-2 \i \pi q} \ 
\ee
for which the partition function becomes a sum over two terms
\be
Z\left[ 0, q \right] = \frac{1}{\sqrt{3}} \left(
e^{2\ell(1-q+i \sigma_+)} -
e^{2 \ell(1-q+i \sigma_-) } \right) \ .
\ee
This expression has a vanishing derivative with respect to $q$ 
and can be simplified to give the following constant 
\be
\label{specialcase}
Z\left[ 0, q \right] = \frac{2}{\sqrt{3}} \sinh \left[ \frac{\Li_2( e^{\i \pi/3} ) - \Li_2( e^{-\i \pi/3} )}{2 \pi \i} \right] \approx 0.379568 \ .
\ee
The argument of the hyperbolic sine function is the volume of an ideal tetrahedron in ${\mathbb H}^3$, $1.01494\ldots$, divided by $\pi$.  The result (\ref{specialcase}) is a special case of the general integral formula proven in ref.\ \cite{Garoufalidis:2014ifa}.
The independence of the result with respect to $q$ follows from the possibility of changing the integration variable $\sigma \to \sigma - i q$ and then shifting the contour in (\ref{Z}), without crossing a pole.  

We can analyze $\tau=1$, for which there are two poles as well, but we shall go no higher to avoid solving a cubic polynomial.
In this case, the poles are at
\be
x_\pm = \frac{e^{-\i \pi q}}{e^{\i \pi q} \pm 1}
\ee
and the partition function evaluates to
\be
Z[1,q] = \frac{\i}{2} \left( \frac{e^{\i \pi \sigma_+^2 +2 \ell(1-q+i \sigma_+))} }{1 + e^{\i \pi q}} 
+ \frac{e^{\i \pi \sigma_-^2+2 \ell(1-q+i \sigma_-)} }{1-e^{\i \pi q}}  \right) \ .
\ee
Through the use of dilogarithm identities, this expression can be massaged somewhat.  Note the first term is close to the square of (\ref{Zhalf}).  The same sort of techniques that yield (\ref{Zhalftwo}) from (\ref{Zhalf}) can be used on the second term, and the end result is
\be
Z[1,q] =\frac{1}{2} 
 e^{- \frac{\i \pi}{12}(3q-2)^2 + \frac{2 \i \pi}{3} }
 \left( e^{2 \ell \left(\frac{q}{2}\right) }- 2 \cos \left(\frac{\pi q}{2} \right) e^{2 \ell \left(\frac{q+1}{2}\right) + \frac{\i \pi}{4}}  \right) \ .
\ee
This partition function has minima at $q = 0.231071$ and $q = 1.44196$.  If we allow the two chiral fields to have different R-charge, however, we find that in the full flavor space, $q = 1.44196$ is a saddle point while $q = 0.231071$ remains a minimum.

\subsubsection*{$n_+ = 1$ and $n_- = 1$}

For $\tau = r$ an integer, the locations of the poles are given by solutions of the polynomial equation
\be
x^{1+r} - e^{2 \i \pi q} x^r - e^{\i \pi(2q+r)} x + e^{\i \pi r} = 0 \ .
\ee
When $r = 0$,
there is a single pole at $ z = \frac{\i}{2}$, and the partition function evaluates to 
\be
\label{XYZcase}
Z\left[ 0 , q \right] = \frac{ e^{2 \ell \left( \frac{1}{2} - q \right)} }{2 \sin(\pi q) } \ .
\ee
This case was analyzed already by ref.\  \cite{Jafferis:2010un} in the context of this model's
duality to the XYZ model -- a theory with three hypermultiplets $X$, $Y$, and $Z$, a superpotential, and no gauge field.  
The relation can be made more explicit at the level of the partition function by employing a dilogarithm
identity to rewrite the right hand side of (\ref{XYZcase}) as $e^{2 \ell(q) - \ell(2q-1)}$.   
Ref.\  \cite{Jafferis:2010un} also noted 
the extremum at $q= \frac{1}{3}$, in correspondence with the fact that the scalar $X$ field can be thought of as a composite 
of two fermions in the gauge theory.
There are other special values of $q$.
At $q = 0$, 1, and 2, $\operatorname{Re} (\log Z) \to \infty$ while at $q = 3/2$, $\operatorname{Re}(\log Z ) \to -\infty$.
The physical reasons behind these divergences remain obscure to us.

When $\tau = 1$,
there are poles at $z = \frac{\i}{2}$ and $z = 0$.  (The pole at $z=0$ can be swapped for one at $z = \i$.)  
\be
Z[1,q] = \frac{ \cot(\pi q) - \i}{4} e^{2 \ell(-q)} + \frac{1}{2} (-1)^{3/4} e^{-\i \pi q} e^{2 \ell\left(\frac{1}{2} - q\right)} \ .
\ee
Numerically, this function has a minimum at $q = 0.384466$.  There is also a local minimum at $q = 1.61553$.
However, if we allow the charges to shift in opposite directions as well, $q_+ = q + \delta q$ and $q_- = q - \delta q$,
then we see that the critical point at $q = 1.61553$ is only a saddle point while $q = 0.384466$ remains a minimum.

\subsubsection*{$n_+ = 2$ and $n_- = 2$}

As the number of flavors increases, the polynomial determining the location of the poles becomes more complicated.
In this case, we have
\be
x^{r+2} - 2 e^{2 \i \pi q} x^{1+r} + e^{4 \i \pi q} x^4 - e^{\i \pi (4q+r)} x^2 + 2 e^{\i \pi (2q +r)} - e^{\i \pi r} = 0
\ee
having set again $\tau = r$ to an integer.
When $\tau = 0$, there are two poles, one at $z = \frac{\i}{2}$ and then one that we may either take at $ z = 0$ or $ z = \i$.
The partition function is 
\be
Z[0, q] = \frac{e^{4 \ell \left( - q \right)}}{32 \sin^{3} (\pi q) \cos(\pi q)} 
-\frac{e^{4 \ell \left( \frac{1}{2} - q \right)}}{4 \sin(2\pi q) } \ .
\ee
There is an extremum at $q = 0.408533$.  
There is also singular behavior, $\operatorname{Re} (\log Z)\to \infty$ at $q = 0$, 1, 3/2, and 2, while at 
$q = 1.187$ and 1.713, $\operatorname{Re}( \log Z )\to -\infty$.

\section{S Duality and Transport}
\label{sec:sduality}

In this section, we investigate 
a strong-weak coupling duality $\tau \to -1/\tau$ that we implement at the level of 
our hemisphere partition function by Fourier transform.  We will pay particular attention to two very simple theories,
the $n_+ = n_- = 1$ theory and the $n_+ = 1$, $n_- = 0$ theory.  In both instances, we will be able to identify
a particular value of $\tau$ where the theory becomes almost self-dual.  Inspired by
refs.\ \cite{Herzog:2007ij,Hsiao:2017lch} but with some crucial differences, 
we can use ``self-duality'' to extract the conductivity.  

\subsection{The $n_+ = n_- = 1$ theory}

The trick to implementing the Fourier transform of the partition function is the residue method we described
in the previous section.
For the theory with $n_+ = n_- = 1$, 
the Fourier transform of interest is a generalization of the result (\ref{XYZcase})
of the  $\tau = 0$ partition function: 
\be
\int \d \sigma e^{\ell(1-q_+ + \i \sigma) + \ell(1-q_- - \i \sigma) + 2 \pi \i k \sigma}  = e^{\ell(1-q_+-q_-) + \ell( \frac{q_+ +q_-}{2} + \i k) 
+ \ell(\frac{q_+ +q_- }{2} - \i k) + \pi k (q_+ -q_-) } \ .
\ee
Previously, we used the $k=0$, $q_+ = q_-$ case of this integral to review the relation between the strong coupling limit of the $n_+ = n_- = 1$ theory and the XYZ Wess-Zumino model. 

With this Fourier transform, we can perform an S duality on the partition function for the $n_+ = n_- = 1$ case with arbitrary $\tau$:
\begin{eqnarray}
\label{n1n1duality}
\lefteqn{\int \d \sigma  e^{\i \pi \tau \sigma^2 +\ell(1-q_f - q_g + \i \sigma) + \ell(1-q_f + q_g - \i \sigma) + 2 \pi q_t \sigma} }\\
&=& \frac{e^{\ell(1-2q_f) }}{\sqrt{-\i \tau}} \int \d k \, e^{-\i  \pi k^2 / \tau + \ell( q_f +q_t+ \i k) 
+ \ell(q_f - q_t - \i k) + 2\pi (k - \i q_t) q_g} \ .  \nonumber
\end{eqnarray}
This S duality is a generalization to arbitrary coupling of the XYZ duality.\footnote{%
 Promotion of this type of purely 3d duality to 4d was discussed in a nonsupersymmetric context in \cite{Seiberg:2016gmd}.
}
The duality has swapped the role of the topological and the gauge symmetry; we can introduce primed quantities on the right hand side of the equality such that
$q_g' = -q_t$ and $q_t' = q_g$.
By symmetry, we expect the partition function to be minimized when $q_t = q_g= 0$.  
From the presence of the $\ell(z)$ functions, we can read off that 
the new theory has two chiral multiplets of R-charge $1 - q_f \pm q_t$ which couple to a U(1) gauge field
with strength $-1/\tau$ and a third decoupled chiral with R-charge $2q_f$.  The sum of the R-charges is two, commensurate with having a cubic superpotential linear in each of the three fields.  
In the limit $\tau \to 0$, we can evaluate the integral by saddle point, recovering (\ref{XYZcase}).  
Indeed, one can analyze the system not only at the strong coupling fixed point but near it as well.  
Near the strong coupling fixed point, $|Z|$ is extremized when $q_+ = q_- = q_f$ and
\be
q_f = \frac{1}{3} + \frac{54 \sqrt{3} - 90 \pi + 8 \sqrt{3} \pi^2}{9 (8\pi (3 \sqrt{3} - 2 \pi)-27)} \operatorname{Im} \tau + O(|\tau|^2) \ .
\ee
We can also compute the current two point functions:
\bea
\Pi_{ff} &=& \Big(\frac{16}{3}-\frac{4\sqrt{3}}{\pi}\Big)+
\frac{8(297-102 \sqrt{3} \pi + 32 \pi^2)}{27(3 \sqrt{3}-4 \pi)} \text{Im}\, \tau
+\mathcal O(|\tau|^{2}) \ ,\nn \\
\kappa_{ff} &=&
\frac{2 \pi(27 - 18 \sqrt{3} \pi + 8\pi^2)}{9 (3 \sqrt{3}-4 \pi)} \text{Re}\, \tau
+\mathcal O(|\tau|^{2}) \ , \\
\Pi_{gg} &=&\Pi'_{tt} = \frac{4}{\pi} \text{Im} \, \tau + 
\frac{4 (4 \pi - 3 \sqrt{3}) \cos(2 \alpha)}{9 \pi} | \tau|^2
+ O(|\tau|^3) \ , \nonumber \\
\kappa_{gg} &=& \kappa'_{tt} = \frac{8}{9}(3 \sqrt{3} - 2 \pi) \pi^2 \sin(2 \alpha) |\tau|^2 + O(|\tau|^3) \ .
\eea
Numerically, the expansion for $\Pi_{ff}$ takes the form $3.13 - 2.32 \, \text{Im}(\tau) + \ldots$.  The 3.13 is larger
than $\Pi_{ff} = 2$ (\ref{largetauPi}) we found 
in the free limit.  The sign of the correction is what we anticipated in the discussion pertaining to figure \ref{fig:deltaPiplot}.
We have made manifest a ``weakly coupled'' description.  
Moving away from this description by increasing the gauge coupling initially tends to 
decrease the conductivity because of increased scattering.  We put weakly coupled in scare quotes because the description is only weakly coupled {\it vis a vis} the 4d gauge field.  There remain interactions governed by a superpotential between the three chiral fields.  

\begin{figure}
\begin{center}
a) \includegraphics[width=2.5in]{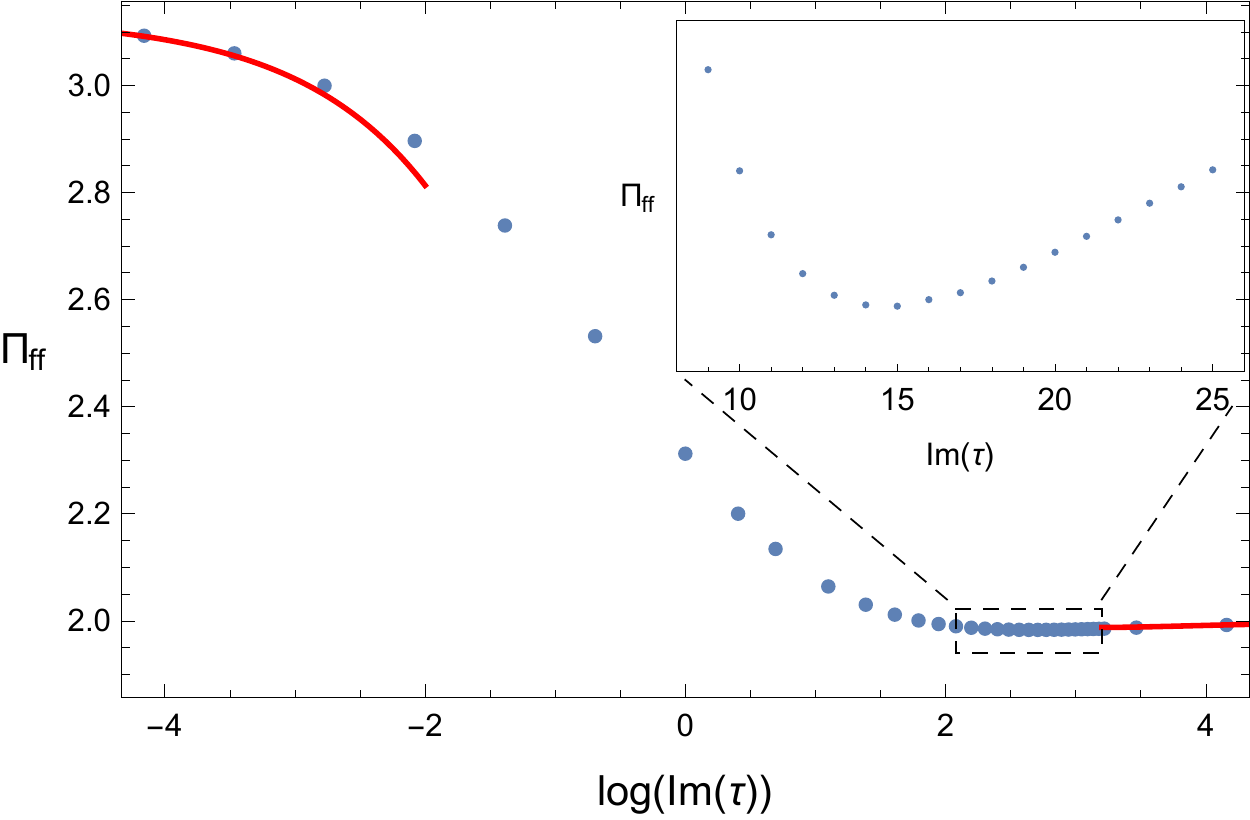}
b) \includegraphics[width=2.5in]{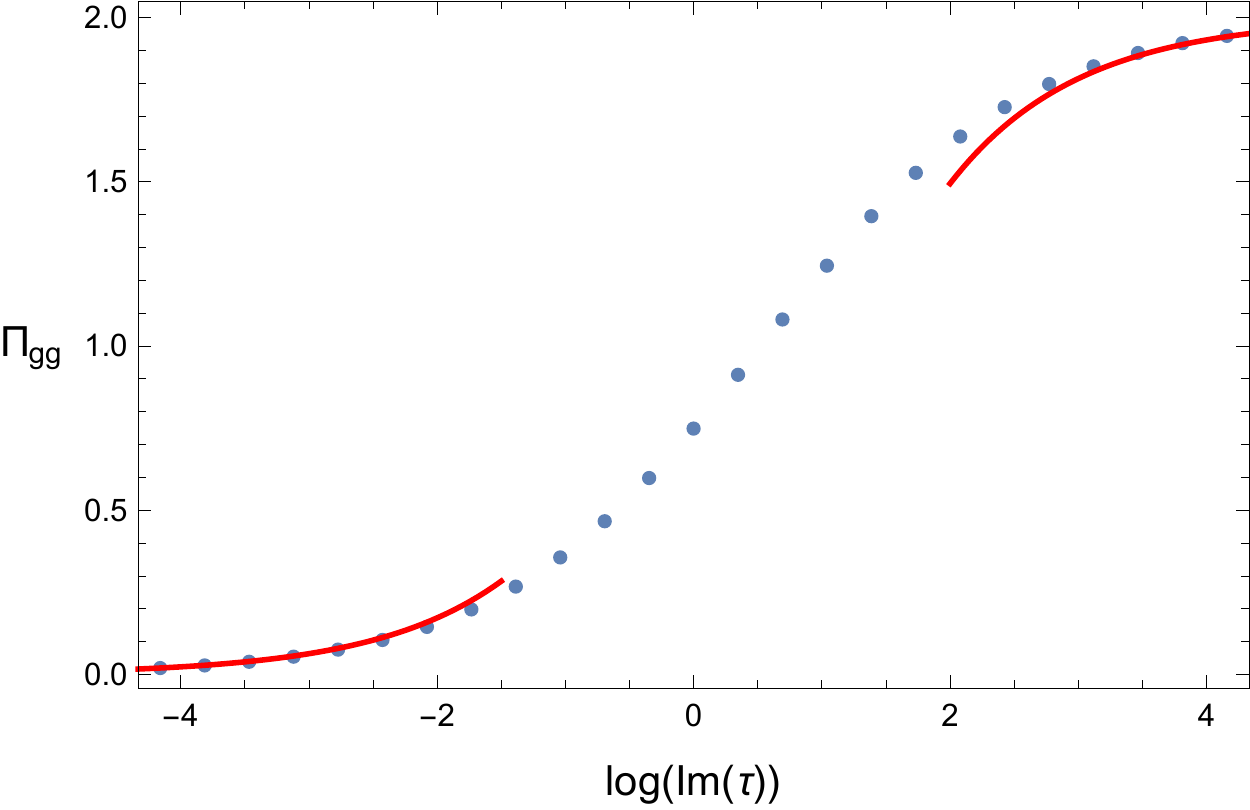}
\end{center}
\caption{A plot of $\Pi$ vs.\ $\text{Im} \, \tau$ for the $n_+ = n_- = 1$ theory.
We set $\text{Re}\, \tau = 0$.  
The solid red curves are saddle point
approximations, while the points are determined from numerical integration.
a) The flavor symmetry.
There is a minimum, visible in the inset, at $\tau \approx 15 \i$.  
b) The gauge symmetry. 
\label{fig:n1n1plot}}
\end{figure}

The full behavior of $\Pi_{ff}$  and $\Pi_{gg}$ as a function of $\text{Im}\, \tau$ for $\text{Re} \, \tau = 0$ can be seen in figure \ref{fig:n1n1plot}.
While $\Pi_{ff}$ has a minimum, the gauged conductivity $\Pi_{gg}$ falls smoothly to zero as $\tau$ is decreased.  
The sum rule (\ref{sumrule}) gives us the behavior of the topological conductivity 
from $\Pi_{gg}$.  
In the strong coupling limit, we find that 
$\Pi_{tt} = \frac{1}{3} \Pi_{ff}$.  Moving away from strong coupling, the topological conductivity decreases, eventually asymptoting to
$\Pi_{tt} = \frac{4}{\pi \text{Im} \, \tau}$.

Something remarkable happens when $\tau = \i$ and $q_+ = q_- = q_f = 1/2$ (and $q_g = q_t = 0$).  The two partition functions on either
side of (\ref{n1n1duality}) become
manifestly equal, up to a factor of $e^{\ell(0)}/ \sqrt{-\i \tau}$, which evaluates anyway to one.  This equality points toward
a self-duality of the underlying gauge theory.  
It is not in fact self-duality.
Critically, the partition function  is minimized not at $q_f=1/2$ but at $q_f = 0.393814$.  There is an additional neutral chiral multiplet in the dual frame that is not present in the original frame, and the charged chiral multiplets have different anomalous dimensions.

We can fix the issue about $q$ by taking advantage of the following dilogarithm identity:
\be
e^{\ell(1-2q)} = e^{\ell(1/2-q) + \ell(1-q)-\ell(q-1/2) - \ell(q) } \ .
\ee
The duality statement can then be improved to 
\begin{eqnarray}
\label{improvedduality}
\lefteqn{ e^{\ell(q_f-1/2) + \ell(q_f) }\int \d \sigma  \, e^{\i \pi \tau \sigma^2 +\ell(1-q_f -q_g+ \i \sigma) + \ell(1-q_f +q_g- \i \sigma) + 2 \pi q_t \sigma} }\\
&=& \frac{e^{\ell(1/2-q_f) + \ell(1-q_f)}}{\sqrt{-\i \tau}} \int \d k \, e^{-\i  \pi k^2 / \tau + \ell(q_f +q_t + \i k) 
+ \ell(q_f - q_t - \i k)  + 2 \pi q_g k - 2 \pi \i q_t q_g} \ .  \nonumber
\end{eqnarray}
Up to the factor $(-\i \tau)^{-1/2}$ suggesting the presence of a decoupled Maxwell field in the dual frame, 
the theory at coupling $\tau$ with R-charges parametrized by $q_f$ and $q_{g}$
has the same partition function as the theory at coupling
$-1/\tau$ and R-charges parametrized by $q_f$ and $q_{t}$.  By a parity argument, 
extremization will set $q_t = q_g = 0$.
Furthermore, now the partition function is minimized at $q_f=1/2$ when $\tau = \i$.  

Let us say a little more about the field content of this modified theory and its dual.
The original theory has neutral chirals $N_1$ and $N_2$ with R-charges $1/2-q_f$ and $1-q_f$ along with oppositely gauge charged chirals $C_\pm$ with R-charge $q_f \pm q_{g}$.  The dual theory on the other hand has neutral chirals $\tilde N_1$ and $\tilde N_2$ 
with R-charges $1/2 + q_f$ and $q_f$ and gauge charged chirals $\tilde C_\pm$ 
with R-charge $1-q_f \mp q_{t}$.  Interestingly, the charges $1/2 \pm q_f$ 
of the $N_1$ and $\tilde N_1$ are in conflict with the unitarity bound 
in one or the other duality frame almost everywhere. 
(Since these are gauge neutral operators, we expect their R-charges  -- and correspondingly their conformal dimensions -- to be greater than or equal to one half.)
The one exception is the self-dual point $q_{f}=1/2$, where we can divide both sides of the partition function by $
e^{\ell(0)}$ and remove these problematic fields from the theory.\footnote{%
 It would be interesting to see if there is some other dilogarithm identity for $e^{\ell(1-2q_f)}$ which can be employed
 which does not give rise to similar unitarity problems.
 }
Meanwhile, we can form gauge invariant mesonic operators $M = C_+ C_-$ and $\tilde M = \tilde C_+ \tilde C_-$ and a superpotential from $N_2^2 M$ or $\tilde N_2^2 \tilde M$, indicating the theory likely has interactions that are separate
from those mediated by the bulk vector multiplet.

\begin{figure}
\begin{center}
 \includegraphics[width=3in]{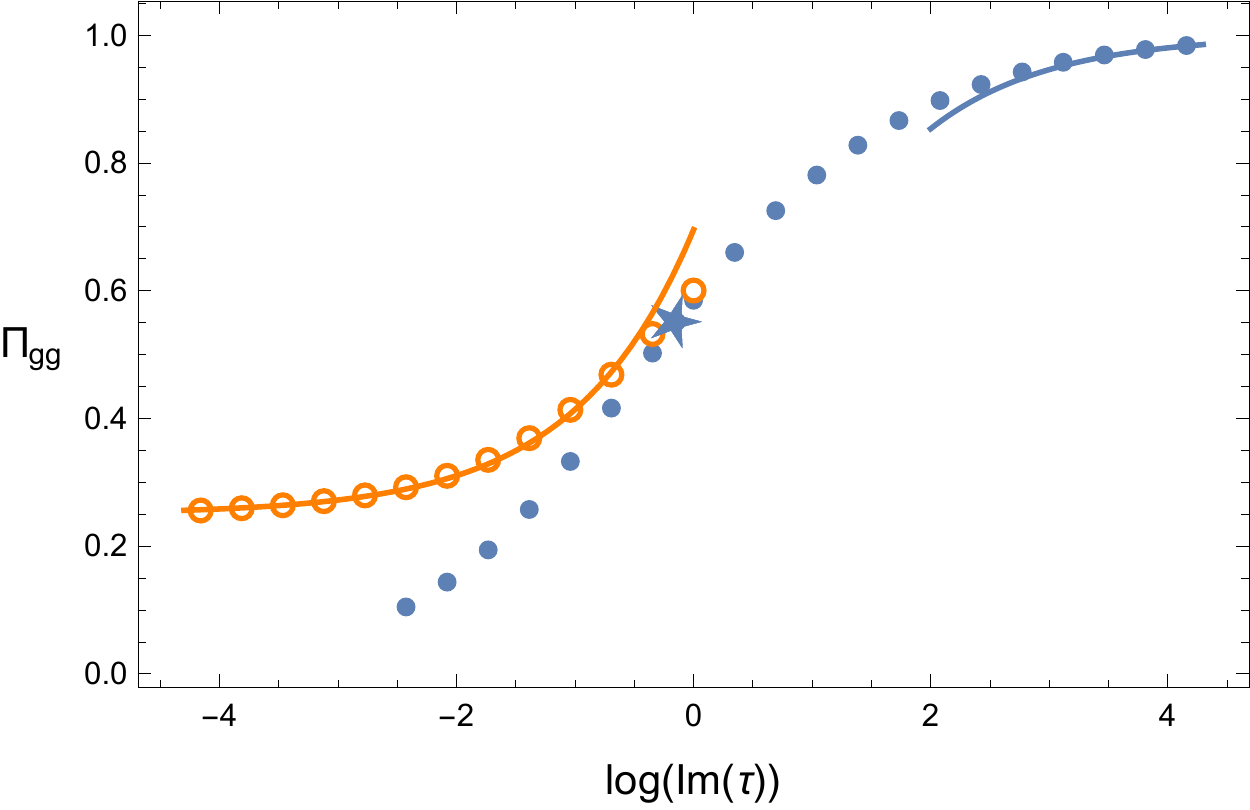}
\end{center}
\caption{A plot of $\Pi_{gg}$ vs.\ $\text{Im} \, \tau$ for the $n_+ = 1$, $n_- = 0$ theory.
We set $\text{Re}\, \tau = 0$.  
The solid points and the star have $\text{Re}\, \tau = 0$ while the open
orange circles have $\text{Re}\, \tau = 1/2$.  The star is the self-dual point.  The curves are 
saddle point approximations.
\label{fig:n1n0plot}}
\end{figure}

Before moving on the to the $n_+ = 1$, $n_- = 0$ theory, we note in passing that 
it is straightforward to generalize this S duality result to the case of $n$ flavors.  We Fourier transform each pair 
of oppositely charged chirals independently, leading to a new theory with $n$ U(1) gauge fields.  For simplicity, we give
the result only in the case when $q_{i+} = q_{i-} = q_i$ where $i = 1, \ldots, n$:
\begin{eqnarray}
\label{Zngeneral}
\lefteqn{\int \d \sigma  e^{\i \pi \tau \sigma^2 +\sum_i \left( \ell(1-q_i + \i \sigma) +  \ell(1-q_i - \i \sigma)\right)} }\\
&=&
\frac{1}{\sqrt{-\i \tau}} \int \d^n k \, e^{-\i  \pi (\sum_i k_i)^2 / \tau + \sum_i \left( \ell( q_i + \i k_i) 
+ \ell(q_i - \i k_i) +\ell(1-2q_i) \right) } \ . \nonumber
\end{eqnarray}
 By Giveon-Kutasov duality \cite{Giveon:2008zn} in the limit $\tau = k$ is an integer, 
 the above result is also supposed to be the partition function of a $U(k +n-1)$ Chern-Simons theory at level $-k$.
 It would be interesting to check that (\ref{Zngeneral}) indeed matches the corresponding localization result
 for this non-abelian theory.

\subsection{The $n_+=1$, $n_- = 0$ theory}

We can play a similar game with the theory of a single chiral multiplet, $n_+ = 1$ and $n_- = 0$, and complexified gauge coupling $\tau$.  
By a change of variables $\sigma \to \sigma + 2 k$, 
we can deduce the Fourier transform of this single chiral theory from the partition function at $\tau = 1/2$, (\ref{Zhalf}):
\be
\int \d \sigma e^{\i \pi \sigma^2 /2 + 2 \i \pi \sigma k   +\ell(1- q_g  + \i \sigma)}  = e^{-2 \i \pi k^2} 
e^{ \ell ( \frac{\tilde q + 1}{2}) 
- \frac{3 \i \pi}{8} \left(\tilde q - \frac{1}{3} \right)^2 + \frac{\i \pi}{12} } \ ,
\ee
where $\tilde q = q_g + 2 \i k$.  This $q$ is most naturally identified with $q_g$ in the theories
with $n_+ = n_-$.
Inserting this Fourier transform into the partition function at arbitrary $\tau$, 
we can perform an S duality and deduce the partition function for
an associated theory with complexified gauge coupling $\tau' = - (\tau - 1/2)^{-1} - 1/2$:  
\begin{eqnarray}
\label{Znonezero}
\lefteqn{\int \d \sigma \, e^{\i \pi \tau \sigma^2 + \ell(1-q_g+\i \sigma) + 2 \pi q_t \sigma} } \\
&=& \frac{e^{- \frac{3 \i \pi}{8} \left(q_g - \frac{1}{3} \right)^2   
+ \frac{\i \pi}{2} q_t^2 
- \frac{3 \i \pi}{2}\left( q_g - \frac{1}{3} \right) q_t 
+\frac{\i \pi}{12}}}{\sqrt{ -\i (\tau - 1/2)}} \int \d k \, e^{\i \pi \tau' k^2  + \ell\left(\frac{q_g+1}{2} + q_t+ \i k\right) 
+ 2 \pi \left(\frac{3q_g-1}{4} -\frac{q_t}{2} \right) k
} \ . \nonumber
\end{eqnarray}
This partition function is that of a single chiral with R-charge $\frac{1-q_g}{2}-q_{t}$.  
There is also a term in the exponent proportional to $2 \pi \left(\frac{3q_g-1}{4} -\frac{q_t}{2} \right) k$, which corresponds to a monopole operator of dimension $(3q_g-1)/4 - q_t/2$ \cite{Jafferis:2010un}.\footnote{%
 We do not expect to be able to construct BPS operators from this monopole operator, and
 there are no corresponding unitarity bound constraints.
}  

We can use a saddle point approximation to compute the R-charge near $\tau = \frac{1}{2}$.  
The calculation is more subtle than before because $q_t$ is nonzero near $\tau = \frac{1}{2}$.  We find that
\bea
\label{qnonezero}
q_g &=& \frac{1}{4} + O(|\tau'|^{-1}) \ , \; \; \; 
q_g + 2 q_t = \frac{2 \sin(\alpha')}{\pi |\tau'|} + O(|\tau'|^{-2})  \ ,
\eea
where $\tau' = |\tau'| e^{\i \alpha'}$.  
The combination $q_g + 2 q_t$ appears because in the limit $\tau \to 1/2$ where the complexified
combination $q_t + \tau q$ becomes real and the two charges can no longer be distinguished,
as was discussed above.
The saddle point approximation also gives the two-point function of the symmetry current in this limit:
\bea
&& \Pi_{gg} = \frac{1}{4} + \frac{ \left( \frac{52}{\pi} - 3 \pi\right) \sin(\alpha') - 12 \cos(\alpha')}{16 |\tau'|} + O(|\tau'|^{-2} \ , \nn \\
&& \kappa_{gg} = \frac{3}{8} + 
\frac{3((\pi^2 - 12) \cos(\alpha') - 4 \pi \sin(\alpha'))}{64 |\tau'|} + 
O(|\tau'|^{-2}) \ .
\eea

We did not analyze this model via saddle point in the weak coupling $\tau \to \infty$ limit before, but let us do so here. 
The weak coupling limit is
\begin{eqnarray}
q_g &=& \frac{1}{2} - \frac{\sin(\alpha)}{\pi |\tau|}  
- \frac{\pi^2 -4 + 2 (\pi^2 + 2) \cos(2 \alpha)}{4 \pi^2 |\tau|^2} 
+ {\mathcal O}(|\tau|^{-3})
\ ,  \nonumber \\
q_t &=& - \frac{\cos(\alpha)}{4 |\tau|} + O(|\tau|^{-2}) \ , \nonumber  \\
\Pi_{gg} &=& 1 - \frac{3 \pi^2 - 16}{4 \pi |\tau|}\sin(\alpha) + {\mathcal O}(|\tau|^{-2}) \ , \nonumber \\
\kappa_{gg} &=& \frac{3 \pi^2 \cos(\alpha)}{16 |\tau|} 
+ {\mathcal O}(|\tau|^{-2}) \ . 
\end{eqnarray}
The characteristics of $\Pi_{gg}$ as a function of $\tau$ are similar to what we found for 
$\Pi_{gg}$ for the $n_+ = n_- = 1$ theory.
As we increase the coupling or equivalently decrease $\tau$, $\Pi_{gg}$ decreases because of the increased scattering.
We have plotted $\Pi_{gg}$ vs.\ $\tau$ in figure \ref{fig:n1n0plot}, demonstrating that the saddle point approximations work reasonably well.

Something remarkable happens when $\tau$ takes its fixed point value $\i \sqrt{3}/2$ under the TST transformation
$-(\tau-1/2)^{-1/2} - 1/2$. The extremal value of the R-charge is $q=1/3$. 
At this point, $e^{\i \pi / 12}$ in the numerator cancels against the $\sqrt{-\i (\tau - 1/2)}$ in the denominator, 
and the two sides of   (\ref{Znonezero}) become manifestly equal.
We view this equality as evidence for an underlying self-duality of the physical theory.
Unlike the $n_+ = n_- = 1$ case, we have no need to invoke dilogarithm identities or modify the original setup.

\subsection{Transport}

The conductivities of 
both the modified $n_+ = n_- = 1$ theory and the $n_+ = 1$, $n_- = 0$ theory take a particularly simple form at the self-dual point, namely
\be
\label{Sigmaggsd}
\Sigma_{gg} = \frac{\tau}{2} \ .
\ee
We would like to try to explain this fact by digging a little deeper into the field theory relations suggested by the integral
equalities
(\ref{improvedduality})
and
(\ref{Znonezero}).
In both cases, we can identify a doublet of transformed charges $(q'_g, q'_t)$ which are related to the original charges 
$(q_g, q_t)$ by a certain affine transformation.  In the $n_+ = n_- = 1$ case, $q_g' = -q_t$ and $q_t' = q_g$, while for the 
$n_+ =1$, $n_- = 0$ case we find the more involved 
$q_g' = \frac{1}{2} - \frac{1}{2} q_g - q_t$ and 
$q_t' = -\frac{1}{4} + \frac{3}{4} q_g - \frac{1}{2} q_t$.  From the discussion around the definition of the current-current two-point functions (\ref{Pikappadef}), it is clear that single derivatives of the partition function with respect to $q_t$ and $q_g$ are related to the associated currents $J_t$ and $J_g$, from which we may deduce the following transformation rule on the doublet 
\be
\label{Jtransform}
\left( \begin{array}{c}
J_g' \\
J_t' 
\end{array}
\right)
= \left(
\begin{array}{cc}
a & b \\
c & d 
\end{array}
\right)
\left( \begin{array}{c}
J_g \\
J_t 
\end{array}
\right)
\ee
implied by the integral relations (\ref{improvedduality})
and
(\ref{Znonezero}).\footnote{%
These rules were also found in \cite{DiPietro:2019hqe}.
}  The element of $SL(2, {\mathbb R})$ involved is the same one that transforms $\tau$ to the $\tau'$ of the dual theory.

%
%
%
%
%
%

This transformation rule (\ref{Jtransform}) implies a corresponding rule on the coefficients $\Sigma_{gg}$, $\Sigma_{gt}$, and $\Sigma_{tt}$ of the current two-point functions.\footnote{%
 We do not actually need the intermediate step of introducing a transformation rule on the doublet $(J_g, J_t)$.  
 We can use the definition of $\Sigma_{ij}$ in the text below (\ref{Pikappadef}) to deduce $\Sigma_{ij}'$ from $q_g'$ and $q_t'$.
}
  This rule can then be substantially simplified with the help of the sum rules (\ref{sumrule}), yielding for $\Sigma_{gg}$ the constraint
\be
\Sigma'_{gg} = \left( a^2 + \frac{2ab}{\tau} + \frac{b^2}{\tau^2} \right) \Sigma_{gg} - \frac{b^2}{\tau} \ .
\ee

Now we impose self-duality.  Staring at the integral relations (\ref{improvedduality})
and
(\ref{Znonezero}), the claim is that $\Sigma'$ should be equal to $\Sigma$ up to the effect of the 
phases that have the schematic form
\be
\label{contact}
e^{  \i \pi A (q_g - q_{0g})^2 + \i \pi B (q_t - q_{t0})^2 +  \i \pi C (q_t - q_{t0}) (q_g - q_{g0})}
\ee
where $A$, $B$, and $C$ are real constants and 
$q_{0g}$ and $q_{0t}$ are the critical values of $q_g$ and $q_t$ at self-duality.  In the particular case of $\Sigma_{gg}$, we find then that
\be
\Sigma_{gg} -A = \left( a^2 + \frac{2ab}{\tau} + \frac{b^2}{\tau^2} \right) \Sigma_{gg} - \frac{b^2}{\tau} \ .
\ee  
For the $n_+ = n_- = 1$ theory, $A = 0$, $a = 0$ and $b=-1$.  For the $n_+ = 1$, $n_- = 0$ theory, $A = -\frac{3}{8}$, 
$a = -\frac{1}{2}$, and $b = -\frac{3}{4}$.  In both cases, one finds (\ref{Sigmaggsd}).

This self-duality, with its additional phases (\ref{contact}), is more delicate than what is usually discussed.
 The idea of a complexified conductivity transforming under $SL(2, {\mathbb Z})$ has a long history in the condensed
 matter literature, going back at least to \cite{Lutken:1991jk}.  In Witten's  \cite{Witten:2003ya}  large $n$ set up for instance, $\tau$ and $\Sigma_{gg}$ transform in precisely the same way under $SL(2, {\mathbb Z})$, suggesting that at a self-dual fixed point, $\Sigma_{gg}$ should be equal to $\tau$, not $\tau/2$.

Unlike graphene, our theory has an exact 2+1 dimensional Lorentz invariance where both the photon and electron travel  at the same speed.  
In this context, refs.\ \cite{Hartnoll:2007ip,Herzog:2009xv} 
pointed out that Ward identities relate other transport coefficients to the conductivity.  
In particular, given the conductivity,
one can also compute the thermoelectric coefficient and the heat conductivity.
  The arguments of refs.\ \cite{Hartnoll:2007ip,Herzog:2009xv} should be revisited in the present case, as they
  rely on having a 2+1 dimensional rather than a 3+1 dimensional stress tensor.  We are optimistic that the conclusions
  are unchanged.
 In a nonsupersymmetric version of our
theory,
refs.\ \cite{Hsiao:2017lch,Hsiao:2018fsc} in fact use self-duality to compute these coefficients directly.\footnote{%
 See also \cite{Hartnoll:2007ih} for a hydrodynamic perspective.
}

In standard discussions of self-duality, 
there is an interesting generalization to nonzero temperature and oscillating electric fields.  
In general, the optical conductivity 
may depend on a function of $\omega/T$ where $\omega$ is the frequency of the applied electric 
field \cite{Damle:1997rxu}.  
Self-duality is supposed to constrain the conductivity to be independent of $\omega$ \cite{Herzog:2007ij}.  It would 
be interesting to explore if those arguments apply to the present case.

Finally, one can explore the consequences at nonzero charge density and magnetic field, which are relevant for
quantum Hall physics.
The boundary condition on the field strength 
relates the charge density $\rho$ to the boundary value of the electric field normal to the surface, and S duality will swap this electric field with the magnetic field $B$.  Refs.\ \cite{Hsiao:2017lch,Hsiao:2018fsc} already
exploited this feature to draw some conclusions about the quantum Hall effect at 1/2 filling in the nonsupersymmetric context.  For carefully chosen values of $\rho$ and $B$, the theory will be invariant under S duality up to a relative
change in sign between $\rho$ and $B$ which may in turn send $\sigma_H \to - \sigma_H$ if $\alpha = \pi/2$.  With this constraint,
one can fix $\sigma$ and $\sigma_H$.  We will leave a  more detailed
investigation of these and other transport issues to future work.

\section{Discussion}
\label{sec:discussion}

We had initially three different motivations for this project: understanding quantities that decrease under renormalization group flow in quantum field theory; the possibility of spontaneous flavor symmetry breaking in three dimensional QED; and boundary conditions in maximally supersymmetric Yang-Mills theory in four dimensions.  While we laid some groundwork in each of these three directions, the main progress came from an unexpected direction -- transport and duality.  We were able to calculate conductivities at all values of the coupling in our theories.  
Furthermore, a pair of our theories turned out to be self-dual.

In conclusion, however, let us return to the original motivation.
The best known example of a quantity that decreases under renormalization group flow is the $a$-anomaly.  In conformal field theory, scale invariance implies that the trace of the stress tensor vanishes classically, $T^\mu_{\; \mu} = 0$.  However, quantum effects on a curved manifold lead to anomalous contributions which are proportional to curvature invariants.  The $a$-anomaly, present in even dimensional cases, is the coefficient of the Euler density term, $T^\mu_{\; \mu} = a E_d + \ldots$, where the ellipses denote other possible curvature invariants.  
Regarding conformal field theories as fixed points in the renormalization group flow, it is proven that $a_{\rm UV} > a_{\rm IR}$ in $d=2$ and 4 \cite{Zamolodchikov:1986gt,Komargodski:2011vj}.  
The $a$-anomaly can also be defined using the sphere partition function; in particular, in even dimensions, there is an anomalous contribution $\log Z = 2 a \log \Lambda + \ldots$ where $\Lambda$ is an energy cut-off.  
While $\log Z$ is finite in odd dimensions, this relation to sphere partition functions suggests that more generally 
there may be a special role for $\log Z$ as a renormalization group monotone.  Indeed, 
in three dimensions, it is also proven using entanglement entropy techniques 
that $F = - \log Z$ decreases under renormalization group flow \cite{Casini:2012ei}.

Of course, our example is not a sphere but a hemisphere.  
The candidate renormalization group monotone for the four dimensional hemisphere is $F_\partial$ 
(\ref{Fpartial}) \cite{Gaiotto:2014gha}.
As we discussed, up to some $\tau$ and $\bar \tau$ dependence that can be fixed by considering the free limit, 
our $\log Z$ is closely related to $F_\partial$.
 
A nice feature of these renormalization group monotones is that they are independent of certain marginal couplings.
The quantity $a$ is independent because of Wess-Zumino consistency \cite{Wess:1971yu,Osborn:1991gm}.  
Ref.\ \cite{Gerchkovitz:2014gta} used conformal symmetry to show $F$ is independent as well.
However, there is a loophole concerning $F_{\partial}$, as one of us showed in 
ref.\ \cite{Herzog:2019rke,Bianchi:2019umv}.  While $F_{\partial}$ is not expected to depend on boundary marginal deformations, it can depend on bulk ones.
A motivation of this work was to explore precisely how $F_{\partial}$, or equivalently $Z_{HS^4}$, can depend on a
bulk marginal parameter, namely $\tau$.  

Looking to the future, there are two anomalous contributions to the stress tensor trace which are associated purely with the boundary, one proportional to the extrinsic curvature cubed and one proportional to a product of the extrinsic curvature and the Weyl curvature \cite{Herzog:2015ioa}.  
It would be interesting to see whether these contributions can be isolated by generalizing
our result to the case of a squashed hemisphere.  The localization result for squashed three and four spheres is already known \cite{Hama:2012bg,Hama:2011ea}.  
 
 Regarding spontaneous symmetry breaking, an interesting idea put forward already in early work \cite{Gorbar:2001qt}
 on these graphene-like theories,
  is that, at least in the non-supersymmetric case with $n$ Dirac boundary fermions, 
for large enough
$g$, the global flavor symmetry may spontaneously break $U(2n) \to U(n) \times U(n)$ 
and a mass gap for the fermions appear.  Part of the motivation of this work was to see if the behavior of the partition function makes such an effect visible in a supersymmetric example. 
The short answer is no, but let us follow this train of thought a little further.

A similar spontaneous breaking of flavor symmetry may also play a role in three dimensional QED \cite{Pisarski:1984dj,Appelquist:1986fd}.  While the gauge coupling becomes large at low energy, there is a ``line''
of fixed points associated with the number $n$ of fermions.  It is conjectured that below a critical value of $n$,
this purely three dimensional theory may also generate a mass gap.  The similarity between the graphene-like theory and ordinary three dimensional QED is no accident \cite{Kotikov:2016yrn}.  Perturbatively in $g$ and at large $n$, 
the Feynman rules of one theory can be mapped onto the other with the substitution $g^2 \sim 1/n$.  
This map means that establishing a critical value for $g$ in our theory may shed light on the critical $n$ in three dimensional QED.

One might hope to investigate spontaneous symmetry breaking by looking at R-charge assignments.
The R-charges obtained from the partition function 
can be used to establish the scaling dimensions of certain protected operators.  
If any of these operators fall below the unitarity bound, the theory may become unstable although the 
picture here is not crystal clear.  In the context of renormalization group flows, 
these operators can also decouple and become free with limited effects on the rest of the theory.  
Decoupling has been observed in 
4d and 3d supersymmetric gauge theory examples
 \cite{Safdi:2012re,Agarwal:2012wd,Morita:2011cs,Kutasov:2003iy,Barnes:2004jj}.  
 R-charge assignments thus provide a suggestive if not conclusive way of looking at the stability 
 of the theory.

In our case, candidate theories to look into these effects are those with the smallest number of flavors, i.e.\
the $n_+ = n_- = 1$ and $n_+ = 1$, $n_- = 0$ theories, in the limit $\operatorname{Im}(\tau) \to 0$.  
We found no evidence in either case to suggest an instability.
As is well known, 
the $n_+ = n_- = 1$ theory is dual to the XYZ model as $\tau \to 0$, and this model is expected to be well behaved.

Interestingly, our self-dual modification to the $n_+ = n_- = 1$ theory did have unitarity bound issues.  
The necessity to add neutral scalars $N_1$ and $\tilde N_1$ with R-charges $1/2 \mp q$ meant the theory 
could only be defined at the self-dual point where these scalars could be removed.  Indeed, there is a conflict
in defining a weakly coupled theory with a weak-strong duality that is unstable at strong coupling.  
The weak-strong coupling duality implies then that the theory is also unstable at weak coupling, that it may in fact
only be defined at order one values of the coupling.  
Our modified $n_+ = n_- = 1$ theory resolves this issue by isolating itself and having a good unitary definition only at the self dual point, $\tau = \i$.

The $n_+ = 1$, $n_- = 0$ theory was dual to a free chiral multiplet at $\tau = 1/2$.  
While $\tau=1/2$ is well defined, we have a lingering suspicion that there may be issues
for this theory for $\tau \neq 1/2$ but close to the real line.
We were able to explore
the theory near $\tau = 1/2$ using the saddle point approximation but saw no smoking gun for an instability.
We also found that the $n_+ =1$, $n_- = 0$ theory had a self-dual point at $\tau = \i \sqrt{3}/2$.
Despite not seeing evidence for instabilities in this theory or the unmodified $n_+ = n_- = 1$ theory, these examples
 with small numbers of flavors have a number of interesting features which are worth further exploring.

A useful tool in the decoupling limit is the Witten index for supersymmetric Abelian Chern-Simons theory.  
Using the results of \cite{Intriligator:2013lca}, the Witten index is always greater than or equal to zero for our Abelian theories, forbidding the breaking of supersymmetry, and presumably also flavor symmetry, 
at least in the decoupling limits.

The third motivation was the hope that the work here lays the groundwork for performing a similar localization calculation
of ${\mathcal N}=4$ super Yang-Mills on $HS^4$, coupled to charged matter on the boundary.
Much is known about supersymmetry preserving boundary conditions in ${\mathcal N}=4$ SYM \cite{Gaiotto:2008sa}.
Like the U(1) theory considered here, ${\mathcal N}=4$ super Yang-Mills with a boundary 
 is an example of a boundary conformal field theory with an exactly marginal coupling --
 again the complexified gauge coupling $\tau$.

Finally, we mention in passing that 
the data we obtain here may be useful in applying the bootstrap program to boundary CFT.
In particular, we may be able to constrain a bootstrap of this graphene-like theory or even 
3d QED \cite{Chester:2016wrc}.

\section*{Acknowledgments}
We thank D.~Anninos, C.~Closset, V.~Forini, D.~Gaiotto, S.~Hartnoll, H.~Kim, I.~Klebanov, E.~Lauria, M.~Marino, and L.~di Pietro for discussion.
We would like especially to 
thank A.~Cabo Bizet and I.~Shamir for discussion and collaboration during the early stages of this work.
This research was supported in part by the U.K.\ Science \& Technology Facilities Council Grant ST/P000258/1.
C.H.\ would like to acknowledge a Wolfson Fellowship from the Royal Society.

\appendix

\section{Fermion Conventions}
\label{app:conventions}

The Clifford algebra in Euclidean signature comes from quotienting the free algebra 
associated to the gamma matrices by the relation
\be
\{\gamma^a, \gamma^b \} = 2 \delta^{ab} \ .
\ee
Our gamma matrices are Hermitian $(\gamma^a)^\dagger = \gamma^a$.
Our basis is the same that Narain et al.\ \cite{Gava:2016oep}  use
\be
\gamma^a = \left( \begin{array}{cc}
0 & -\i \sigma^a \\
\i \sigma^a & 0 
\end{array}
\right) \ ; \; \; \;
\gamma^4 =  \left( \begin{array}{cc}
0 & 1 \\
1 & 0 
\end{array}
\right) \ , \; \; \;
\gamma^5 =  \gamma^1 \gamma^2 \gamma^3 \gamma^4 = \left( \begin{array}{cc}
1 & 0 \\
0 & -1 
\end{array}
\right) \ .
\ee

We can define two candidate charge conjugation matrices $C = \gamma_1 \gamma_3 = {\rm diag}(-\i \sigma_2, -\i \sigma_2)$ and 
$\tilde C = \gamma_2 \gamma_4 \gamma_5 = {\rm diag}(-\i \sigma_2, \i \sigma_2)$.  They are both incompatible with a Majorana condition since  $C^* C = \tilde C^* \tilde C = -1$ but compatible with a symplectic Majorana condition.  We choose $C$ to define symplectic Majorana fermions
\be
\bar \psi_i \equiv (\psi_i)^\dagger = \epsilon_{ij} (\psi^j)^T C \ .
\ee
In particular, $\psi_1^\dagger = \psi^{2T} C$ and $\psi_2^\dagger = -\psi^{1T} C$.  Furthermore, $\psi^2 = C \psi_1^*$.  
The matrix $C$ satisfies the further conjugation properties
\be
\gamma_a^* = \gamma_a^T = C \gamma_a C^{-1} \ , \; \; \; C^\dagger = C^T = C^{-1} = -C \ , \; \; \; C^* = C \ .
\ee

\subsection{Curved Space}

In curved space, we define vielbeins ${e^a}_\mu$ such that the combination 
$\delta_{ab} {e^a}_\mu {e^b}_\nu = g_{\mu\nu}$ yields the metric.  
The curved space gamma matrices are then
\be
\gamma^\mu = {e_a}^\mu \gamma^a \ ,
\ee
such that
\be
\{ \gamma^\mu, \gamma^\nu \} = 2 g^{\mu\nu} \ .
\ee
The covariant derivative on the fermion is 
\be
\D_\mu \psi = \partial_\mu \psi - \frac{\i}{4} \omega_{ab \mu} \sigma^{ab} \psi \ ,
\ee
where $\sigma^{ab}$ is a generator of the local Lorentz group in the spinor representation:
\be
\sigma^{ab} = \frac{\i}{2} [ \gamma^a, \gamma^b ]  = \i \gamma^{ab} \ .
\ee

\section{Special Cases of the Localization Integral}
\label{app:specialcases}

For the special case $n=1$ and $q=1/2$, we find the following results, that depend on the parity of $r$ and $s$.

\noindent
{\it $r$ and $s$ both odd or both even:}
\be
 \int \frac{e^{\i \pi \frac{r}{s^2} \sigma^2}}{2 \cosh(\pi \sigma)} \d \sigma =  \frac{\i s}{2r} \sum_{n=0}^{r-1} \frac{e^{-\frac{\i \pi n^2}{r}}}{\cos \left(\frac{\pi n s}{r} \right)}
- \frac{\i}{2} \sum_{m=0}^{s-1} \frac{ (-1)^m e^{-\frac{\i \pi r}{s^2} \left(m+\frac{1}{2}\right) \left(m+\frac{1}{2}-s\right)}}{\sin \left(\frac{\pi r \left(m+\frac{1}{2}\right)}{s} \right)} \
\ee

\noindent
{\it $r$ odd/even and $s$ even/odd:}
\be
 \int  \frac{e^{\i \pi \frac{r}{s^2} \sigma^2}}{2 \cosh(\pi \sigma)}\d \sigma = \frac{\i s}{2r} \sum_{n=0}^{r-1} \frac{e^{-\frac{\i \pi \left(n+\frac{1}{2}\right)^2}{ r}}}{\cos \left(\frac{\pi \left(n+\frac{1}{2} \right) s}{r} \right)}
+ \frac{1}{2} \sum_{m=0}^{s-1} \frac{ (-1)^m e^{-\frac{\i \pi r}{s^2} \left(m+\frac{1}{2}\right) \left(m+\frac{1}{2}-s\right)}}{\cos \left(\frac{\pi r \left(m+\frac{1}{2}\right)}{s} \right)} \ 
\ee

Duality results when $q = 1/2$:

\be
\int \frac{e^{\i \pi \tau x^2}}{\cosh(\pi x)} \d x = \frac{1}{\sqrt{-i \tau}} \int \frac{e^{-\i \pi x^2/\tau}}{\cosh(\pi x)} \d x \ ,
\ee

\be
\sqrt{-\i \tau} \int \frac{e^{\i \pi \tau x^2}}{\cosh(\pi x)^n} \d x = \int \frac{e^{- \i(\sum_i x_i)^2 / \tau}}{\prod_{i=1}^n \cosh(\pi x_i)} \d^n x \ ,
\ee

\be
\int \frac{\d x}{\cosh(\pi x)^n} = \frac{\Gamma \left(\frac{n}{2}\right)}{\sqrt{\pi} \Gamma\left(\frac{1+n}{2}\right)} \ .
\ee

\bibliography{supergraphene}
\bibliographystyle{JHEP}

\end{document}